\documentclass[12pt,preprint]{aastex}

\newcommand{\ts}{\thinspace}
\newcommand{\simless}{\mathbin{\lower 3pt\hbox
     {$\rlap{\raise 5pt\hbox{$\char'074$}}\mathchar"7218$}}}
\newcommand{\simgreat}{\mathbin{\lower 3pt\hbox
     {$\rlap{\raise 5pt\hbox{$\char'076$}}\mathchar"7218$}}}

\newcommand{\about}    {$\sim$\ts}
\newcommand{\aboutless}{$\simless$\ts}

\newcommand{\Spitzer}{{\it Spitzer}}

\newcommand{\ISO}{{\it ISO}}

\newcommand{\etal}{\ts et~al.}
\newcommand{\RM}{$R_{24/8}$}

\begin{document}

\title{The Formation and Evolution of Planetary Systems: 
       Description of the Spitzer Legacy Science Database}

\author{John M. Carpenter\altaffilmark{1}}
\author{Jeroen Bouwman\altaffilmark{2}}
\author{Murray D. Silverstone\altaffilmark{3}}
\author{Jinyoung Serena Kim\altaffilmark{4}}
\author{John Stauffer\altaffilmark{5}}
\author{Martin Cohen\altaffilmark{6}}
\author{Dean C. Hines\altaffilmark{7}}
\author{Michael R. Meyer\altaffilmark{4}}
\author{Nathan Crockett\altaffilmark{8}}

\altaffiltext{1}{Department of Astronomy, California Institute of Technology, 
       Mail Code 105-24, 1200 East California Boulevard, Pasadena, CA 91125}
\altaffiltext{2}{Max-Planck-Institut f\"ur Astronomie, D-69117 Heidelberg, Germany}
\altaffiltext{3}{Eureka Scientific Inc, 113 Castlefern Dr., Cary NC 25713}
\altaffiltext{4}{Steward Observatory, The University of Arizona, 933 North Cherry Avenue, Tucson, AZ 85721}
\altaffiltext{5}{Spitzer Science Center, California Institute of Technology, Mail Code 314-6, 1200 East California Boulevard, Pasadena, CA 91125}
\altaffiltext{6}{Radio Astronomy Laboratory, University of California, Berkeley, CA 94720}
\altaffiltext{7}{Space Science Institute, 4750 Walnut Street, Suite 205, Boulder, CO 80301}
\altaffiltext{8}{Department of Astronomy, University of Michigan, Ann Arbor, MI 48109}

\begin{abstract}

We present the science database produced by the Formation and Evolution of
Planetary Systems (FEPS) \Spitzer\ Legacy program. Data reduction and
validation procedures for the IRAC, MIPS, and IRS instruments are described in
detail. We also derive stellar properties for the FEPS sample from available
broad-band photometry and spectral types, and present an algorithm to normalize
Kurucz synthetic spectra to optical and near-infrared photometry. The final
FEPS data products include IRAC and MIPS photometry for each star in the FEPS
sample and calibrated IRS spectra.

\end{abstract}


\section{Introduction}

The Formation and Evolution of Planetary Systems (FEPS) \Spitzer\ Legacy
program \citep{Meyer06} was designed to characterize the evolution of
circumstellar gas and dust around solar-type stars between ages of 3~Myr and
3~Gyr. To achieve these goals, FEPS obtained spectro-photometric observations
with the \Spitzer\ Space Telescope \citep{Werner04} for a sample of 328 stars
\citep[see][for a description of the sample]{Meyer06}. The observing
strategy was to measure the spectral energy distribution (SED) between
wavelengths of 3.6\micron\ and 70\micron\ with IRAC \citep[InfraRed Array
Camera;][]{Fazio04} and MIPS \citep[Multiband Imaging Photometer for
Spitzer;][]{Rieke04} photometry, and between 8 and 35\micron\ with 
low-resolution IRS \citep[Infrared Spectrograph;][]{Houck04} spectra. In 
addition, the FEPS program obtained MIPS~160\micron\ photometry for 80 stars to
search for colder dust, and high-resolution IRS spectra for 33 sources to probe
for circumstellar gas. 

The FEPS team has produced several studies on the incidence of dusty debris
disks around solar type stars, including the discovery of a debris system in
the initial \Spitzer\ observations \citep{Meyer04}, a census of warm debris
\citep{Stauffer05,Silverstone06,Hines06,Meyer08}, the identification of 
Kuiper-belt analogs \citep{Kim05,Hillenbrand08}, and an investigation of debris
disks around stars with known planets \citep{Moro07}. The FEPS team has also
analyzed the processing of dust in optically thick, primordial disks
\citep{Bouwman08}, and has produced a series of papers on the evolution of gas
in solar-type stars \citep{Hollenbach05,Pascucci06,Pascucci07}.

This paper describes the data reduction procedures for IRAC (3.6, 4.5, 
and 8\micron) and MIPS (24 and 70\micron) images and IRS low resolution
spectra obtained by the FEPS program. Data reduction methods for the
MIPS~160\micron\ images and IRS high resolution spectra are discussed in Kim
\etal~(in preparation) and \citet{Pascucci06}, respectively. The adopted
reduction procedures for the IRAC, MIPS~24\micron, MIPS~70\micron, and IRS
observations are presented in \S\S\ref{irac}-\ref{irs}. We also investigate 
the effects of source confusion on the 24\micron\ and 70\micron\ photometry
(\S\ref{confusion}) and the relative calibration accuracy between \Spitzer\ 
instruments (\S\ref{crosscal}). The series of FEPS papers frequently utilized
synthetic spectra derived from Kurucz model atmospheres to infer the presence
of infrared excesses diagnostic of circumstellar dust. In the Appendices,
we describe the data and algorithm used to obtain normalized synthetic spectra 
for individual stars. The primary data products from the FEPS program are a
tabulation of IRAC and MIPS photometry presented in Table~\ref{tbl:phot},
and extracted, calibrated spectra which are available electronically.

\section{IRAC}
\label{irac}

IRAC produces images in 4 channels at wavelengths of 3.6, 4.5, 5.8,
and 8.0\micron\ with bandwidths of
0.75, 1.01, 1.42, and 2.93\micron, respectively \citep{Fazio04}.
The FEPS team obtained IRAC observations for 311 of the 328
stars in the sample. The remaining 17 objects were observed by other Spitzer
programs, including 16 Hyades stars in a Guaranteed Time Observations (GTO)
program led by G. Fazio, and one source (ScoPMS~214) in the Upper Sco OB
Association by the c2d Legacy Program \citep{Evans03}.

FEPS IRAC observations were conducted in sub-array mode with a four-point
dither pattern and the medium dither scale. The locations of the four dither
positions on the array are the same for each source to within the pointing
accuracy of the spacecraft \citep[$1\sigma$ $< 1$\arcsec\ radial;][]{Werner04}. In sub-array mode, each IRAC band is observed
separately where a $32\times32$ pixel section ($39''\times39''$) in a corner of
the $256\times256$ pixel full-array ($5.2'\times5.2'$) is read out at
frame-times of 0.02, 0.10, or 0.40 seconds. At each dither position, 64 images
are taken at the same frame-time for a total of 256 images per band, with the
same frame-time for each band and a given source. The total on-source
integration time per band is then 5.12, 25.6, and 102.4~sec for frame-times of
0.02, 0.10, and 0.40~sec, respectively. The frame-time was selected on a
source-by-source basis to achieve high signal to noise on the stellar
photosphere without saturating the detector. Five FEPS source were observed in
all four IRAC bands for the initial verification observations. The IRAC
5.8\micron\ observations were dropped for the remaining sources since it had
the lowest signal-to-noise ratio (SNR) of the four IRAC bands. The IRAC
5.8\micron\ data are described in \citet{Meyer04} and are not further discussed
here.

Four-band IRAC GTO observations of the 16 Hyades stars and c2d observations 
of ScoPMS~214 were observed in full-array, high-dynamic-range mode where an 
image is obtained with a 0.6~sec frame-time followed by an image with 
a 12~sec frame-time. The Hyades stars were observed at three dither positions,
and ScoPMS~214 at two positions in this manner. The 12~sec frame-time images 
in IRAC 3.6, 4.5, and 5.8\micron\ bands were saturated and not analyzed. 

\subsection{Image Processing}
\label{irac:process}

Analysis was performed on Basic Calibrated Data (BCD) products generated 
by the data reduction pipeline version S13 developed at the \Spitzer\ Science 
Center (SSC). The SSC pipeline removes the electronic bias, subtracts a dark
image, applies a flat field correction, and linearizes the pixel response.
Additional processing on the BCD images was performed by the FEPS team as now
described.

For the sub-array data, cosmic ray hits were identified and flagged by
filtering the sequence of 64 frames at each dither position. At a given pixel,
the median and standard deviation of the 64 frames values were computed from
the median absolute deviation\footnote{The median absolute deviation (MAD) is
defined as MAD = median$_i(|x_i - {\rm median}_j(x_j)|)$ \citep{Hampel74}. The
standard deviation is estimated from the MAD as $\sigma \approx 1.4826\ {\rm
MAD}$.} to reduce sensitivity to outlier pixel values. Any pixels that deviated
from the median by more than $n\sigma$ were flagged, where $n$ was calculated
to correspond to a probability of 10$^{-4}$ that such an outlier pixel could
occur by gaussian noise given $N$ images (nominally, $N$ = 64 and 
$n \approx 4.8$). For 26 frames (or 0.03\% of the data), the rejected pixel
was within the photometric aperture and the entire frame was discarded. The
median and dispersion were recomputed in an iterative fashion until no
additional pixels were flagged. For sources HD~77407 and HD~70516, we removed
all frames at two dither positions where the FEPS target position overlapped
with a latent image. 

For the archival full-array observations, cosmic-ray rejection was performed
by the MOPEX\footnote{http://ssc.spitzer.caltech.edu/postbcd}
\citep{Makovoz05} mosaicking package. Images were
aligned spatially based on the World Coordinate Systems (WCS) parameters
in the image headers. The standard deviation at each pixel position in the
stack of aligned images was computed from the median absolute deviation.
Pixel values that deviated more than 5$\sigma$ from the median were removed.

After outlier rejection, both the sub- and full-array images were multiplied by
the photometric correction images produced by the SSC that account for
variations in the pixel solid angle and the effective response of the filters
across the IRAC focal plane\footnote{The correction images are available at
http://ssc.spitzer.caltech.edu/irac/calib.}. These correction
images were derived by observing a star at 225 positions across the full-array,
and thus link the calibration of sub- and full-array observations. 

\subsection{Photometry}
\label{irac:phot}

IRAC photometry was measured with a modified version of
IDLPHOT\footnote{http://idlastro.gsfc.nasa.gov}. We measured source flux
densities with aperture photometry instead of point-response-function (PRF)
fitting photometry since the PRF is undersampled in the IRAC~3.6\micron\ band
and the PRF shape depends on the position of the star within a pixel. Aperture
photometry was performed on each frame with an aperture radius of 3 pixels.
This aperture size was chosen as a compromise between signal-to-noise (which
empirically was highest for an aperture radius of 2 pixels), and obtaining
accurate calibration between full and subarray observations (which favored
larger apertures to reduce the effects of image distortion). The sky background
was computed in an annulus on the source centroid with an inner radius of
10~pixels and a width of 10~pixels so that the aperture corrections can be
compared directly to values listed in the \Spitzer\ Observing Manual. Pixel 
values in the sky-annulus were sigma-clipped in an iterative fashion with a
clipping threshold of $3\sigma$, where the dispersion in the sky background was
estimated from the median absolute deviation. The sky value was estimated as
the mean of the remaining pixels. For several sources, the signal to noise
ratio was too low to derive an accurate centroid on individual frames. A subset
of frames was then coadded until the formal, internal positional uncertainty
was less than 0.1 pixels. 

In the IRAC 3.6\micron\ band, the measured flux density can vary up to 3.6\%
depending on the distance of the centroid position from the pixel center
\citep{Reach05}, which is defined as the pixel phase ($p$). This dependency may
be caused by nonuniform quantum efficiency across a pixel. The best-fit
correction factor ($f_{\rm phase}$) derived from the FEPS data is $f_{\rm
phase} = 1.0232 - 0.0582p$, which is similar to that obtained by
\citet{Reach05}. A correlation of similar magnitude between intensity and
pixel phase was found for only one of the four dither positions in the
4.5\micron\ band, and none of the dither positions in the 8\micron\ band.
Pixel phase corrections were applied on individual images for the IRAC
3.6\micron\ band only using the above relation.

Aperture corrections are needed to convert the photometry to the fiducial
10-pixel aperture used to calibrate the IRAC instrument \citep{Reach05}. The
multiplicative scaling factor for the 3-pixel aperture was measured from the
FEPS data by computing the ratio of the flux density in a 10-pixel aperture to
that in a 3-pixel aperture. The derived aperture corrections for a 3-pixel wide
aperture are 1.109, 1.110, and 1.200 for IRAC bands 3.6, 4.5, and
8\micron, respectively. These aperture corrections agree with the values listed
on the SSC IRAC Data Handbook\footnote{http://ssc.spitzer.caltech.edu/irac/dh}
to within 0.3\% for the 3.6 and 4.5\micron\ bands, and 1.5\% for the 8\micron\ 
band. The measured aperture corrections for 23
sources deviated by more than 3$\sigma$ from the nominal value. Twenty of these
sources are known from an adaptive optics survey to be multiple systems with a
separation of \aboutless 2\arcsec\ between the primary and secondary components
(S.~Metchev, private communication). The other three sources have not been
observed at high resolution and the multiplicity status is unknown. For these
23 stars, the measured aperture correction at a four pixel radius is within
1.3\% of the nominal correction for each source, and a four pixel
aperture radius was used with aperture corrections of 1.069, 1.079, 
and 1.081 for IRAC bands 3.6, 4.5, and 8\micron, respectively. These
stars are noted in Table~\ref{tbl:phot}.

\begin{figure}
\includegraphics[angle=0,scale=0.8]{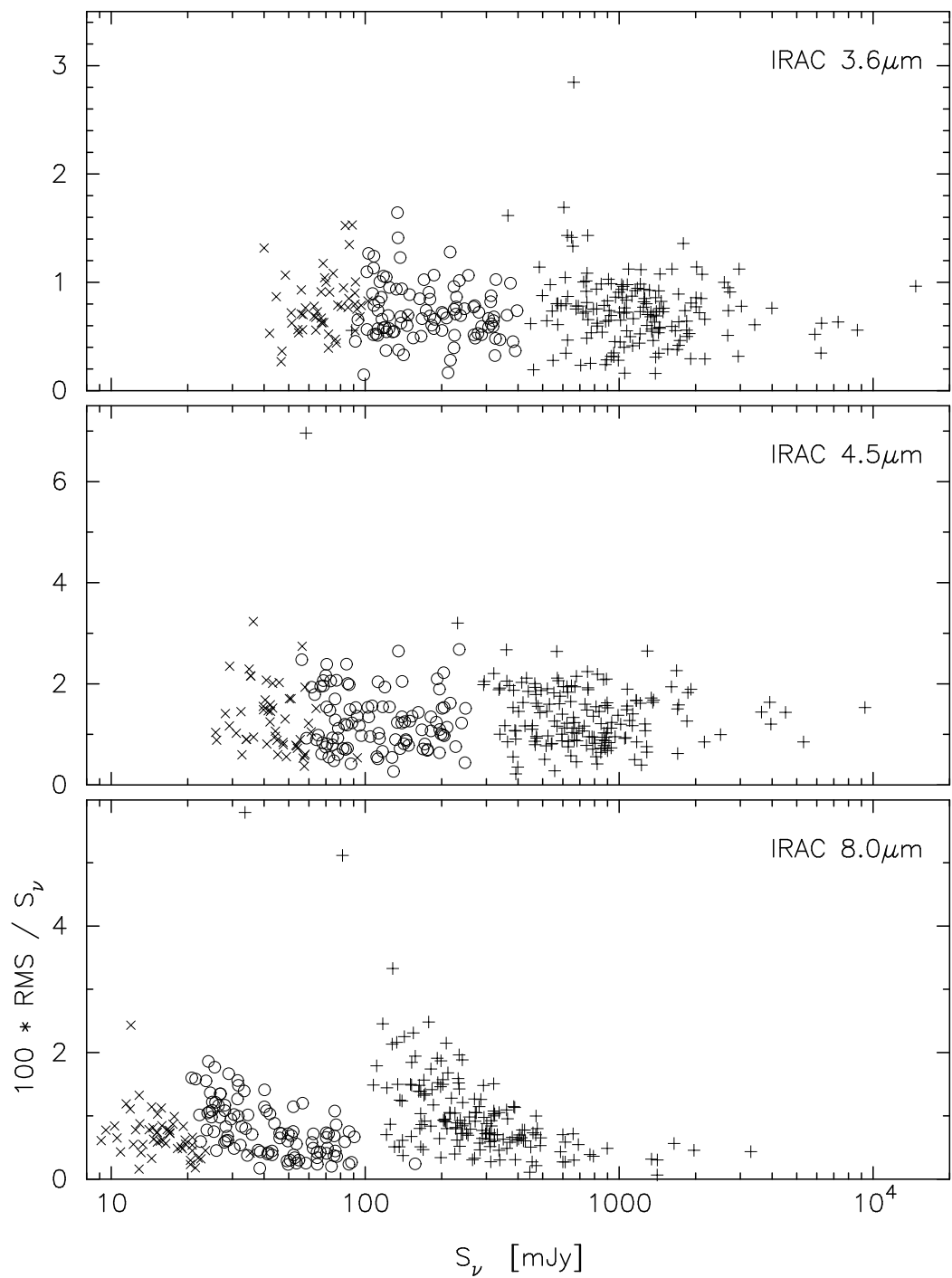}
\caption{
  \label{fig:rms_irac}
  Normalized RMS of the measured flux densities in the four sub-array dither 
  positions plotted versus the mean flux density for the FEPS IRAC sub-array 
  observations. Stars observed with IRAC frame-times of 0.02, 0.1, and 0.4~sec 
  are represented by crosses (+), open circles, and times symbols ($\times$), 
  respectively.
  We used the repeatability between the dithered observations to assign a 
  minimum photometric uncertainty of 0.72\%, 1.22\%, and 0.66\% for IRAC bands 
  3.6, 4.5, and 8\micron, respectively.
}
\end{figure}

Flux densities were computed as the unweighted average of the flux densities
measured in $N$ dither positions ($N$=4 nominally). The standard deviation of 
the $N$ dither positions ($\equiv \sigma_{\rm RMS}$), normalized by the mean
flux density, is plotted versus the mean flux density
in Figure~\ref{fig:rms_irac}. For the IRAC 3.6 and 4.5\micron\ bands, the
normalized RMS shows no trend with mean flux density, while for IRAC
8\micron, the normalized RMS increases systematically toward fainter
sources for a fixed frame-time. This trend is expected if the signal to noise
is photon limited and the integration time is constant since fainter sources
will have lower signal to noise. The photometric repeatability at a fixed 
dither position indicates that the standard deviation of the photometry 
computed from the four dither positions should be $< 0.4$\% in each band for 
the brighter stars. Given that the repeatability between dithers is poorer,
the photometric precision is limited by either our data reduction procedures or
instrumental limitations in obtaining dithered data. Internal photometric
uncertainties were therefore computed as $\sigma_{\rm RMS}/\sqrt{N}$ but with a
minimum uncertainty imposed. For the IRAC 3.6 and 4.5\micron\ bands, we adopt a
minimum internal uncertainty of 0.72\% and 1.22\% respectively, which
corresponds to the median repeatability from the ensemble data shown in
Figure~\ref{fig:rms_irac}. For the IRAC 8\micron\ band, we adopt a minimum
uncertainty of 0.66\%, which is the median value for stars with a repeatability
less than 1.2\%.

We investigated the relative calibration of IRAC sub- and full-array data
since the flux calibrators used by \citet{Reach05} were observed in
full-array mode. To compare the sub-array and full-array calibration, we
analyzed observations of the star HD~135285 that were obtained by the SSC in
full-array mode and in sub-array mode with 0.4~sec integration times. The mean
ratio of the flux densities measured in sub-array mode to that in full-array
mode is $1.004 \pm 0.004$ for the IRAC 3.6\micron\ band, $1.001 \pm 0.004$ for
4.5\micron, $0.995 \pm 0.002$ for 5.8\micron, and $0.997 \pm 0.001$ for
8\micron. The weighted mean for all four bands is $0.997 \pm 0.001$. We
conclude that any calibration offsets between the 0.4~sec sub-array mode and
full-array mode is less than 1\%, and no further calibration corrections were
applied to the sub-array observations. In \S\ref{crosscal}, we consider the
relative sub-array calibration for the different frame-times.

The IRAC photometry and internal uncertainties are presented in
Table~\ref{tbl:phot}. The flux density measurements are tied to the calibration
described in \citet{Reach05} with calibration factors of 0.0188, 0.1388, and
0.2021 MJy/sr per DN/s for IRAC 3.6, 4.5, and 8\micron\ respectively and a
1$\sigma$ uncertainty of 2\%.

\section{MIPS 24\micron}
\label{mips24}

The MIPS instrument obtains images in the 24, 70, and 160\micron\ bands. This
section describes the data reduction procedures for the 24\micron\ band. 
The 70\micron\ data are discussed in \S\ref{mips70}, and analysis of the
160\micron\ data are presented in Kim \etal~(in preparation). The
$128\times128$~pixel MIPS~24\micron\ array images an instantaneous field of 
view of \about 5.4\arcmin$\times$5.4\arcmin\ region with a pixel scale of
$2.5\arcsec\times2.6\arcsec$. The FEPS team obtained MIPS~24\micron\
observations in photometry mode for 323 sources. Data for five stars 
(HD~17925, HD~72905, HD~202917, HD~216803, ScoPMS~214) were extracted from the
\Spitzer\ archive. The exposure time (either 3 or 10~sec) and the number of
dithered images (either 28 or 56) were set to achieve a signal to noise of at
least 30 on the expected stellar photosphere brightness.

\subsection{Image Processing}
\label{mips24:images}

MIPS 24\micron\ images for all but one source were processed with SSC pipeline 
version S13. The star HD~143006 has a flux density of \about 3~Jy at 24\micron,
and S13 data products have an error in the linearity correction for such
brighter sources. For HD~143006 only, we used S16 data products where the
linearity problem was fixed.

Individual BCD images that contain the ``strong'' jailbar effect caused by
bright sources or cosmic rays were removed upon visual inspection. Images were
also removed if cosmic ray hits were found near the expected source position.
These images were identified by performing aperture photometry on individual
BCD images, and finding outlier flux densities or centroid coordinates compared
to the mean that had less than a 10$^{-4}$ chance to have been caused by random
noise.

Once contaminated BCD images were removed, additional processing steps were
performed following the recommendations from the SSC MIPS handbook and the
MIPS instrument team \citep{Engelbracht07}. First, for a given source, 
background levels in individual images were adjusted to a common median 
value using an additive constant. Images for a given source were then median
combined to derive a flat-field image which removes long term gain changes in
the MIPS array. For the median filtering, a 5-pixel radius region centered on
the source position was masked. A 3$\sigma$ clipping algorithm was used to
remove outliers on a pixel-by-pixel basis through the image stack. The
resulting median image was normalized by the median pixel value over the image.
Flat field images were derived only for sources that are not surrounded by
nebulosity. Affected sources were identified from visual
inspection of the image mosaics. If nebulosity is present, a
flat-field image from another FEPS source was used that was a) obtained within
a time interval $\pm$ 1 day, b) had the same exposure time, and c) had the
closest matched background level. If no such image existed, the image nearest
in time with the same exposure time was used. The stability of the flats over
time were assessed by taking the ratio of flats taken on different days. 
Over a $\pm3$~day period, the mean flat field value for the central $5\times5$ 
pixel region of the MIPS~24\micron\ array is repeatable to 1.4\% peak-to-peak
with a dispersion of 0.2\%.

\subsection{Photometry}
\label{mips24:phot}

Photometry was performed with the MOPEX package \citep{Makovoz05}. The BCD
images for a given source were aligned spatially based on
the WCS information in the image headers. Cosmic ray rejection was performed by
removing pixels within the stack that deviated by more than 5$\sigma$ from the
mean. Point sources were identified on a mosaic of the BCD images using a
10$\sigma$ detection threshold. The detection list was modified after visual
inspection of the mosaics to remove spurious sources and to add any sources
missed by the automated detection method. 

PRF fitting photometry was performed with the APEX
module in MOPEX. PRF fitting photometry was chosen over aperture photometry
since the PRF is critically sampled in the MIPS 24\micron\ images and should
provide the optimum signal to noise. The empirical PRF distributed with the
APEX package was fitted to the individual BCD images simultaneously (as opposed
to the mosaicked image) using a fitting area of 21$\times$21 pixels for
most images. A 5$\times$5 pixel fitting area was used for 11 sources that have
spatially variable nebulosity near the point source position. From visual
inspection of the mosaicked images, the PRF from other 24\micron\ sources 
sometimes overlapped with the PRF from the FEPS target. These contaminating 
sources were fitted with a PRF simultaneously with the FEPS target. The free
parameters in the PRF fit include a spatially-constant sky background level,
and the peak flux density and centroid position for each source.

Photometry was measured by integrating the fitted PRF within a 3 pixel radius
(1 pixel \about 2.55\arcsec) since the wings of the PRF have lower signal to
noise. An aperture correction is then needed to place the PRF photometry on 
the zero-point scale adopted by the MIPS instrument team. The aperture 
correction was derived by measuring aperture photometry on individual BCD
images using a customized version of IDLPHOT. We adopted an aperture radius of
13\arcsec\ and a sky annulus between 20\arcsec\ and 32\arcsec\ since these
aperture parameters have been calibrated by the MIPS instrument team to a
theoretical PRF. Aperture flux densities were computed as the unweighted
mean of the photometry measured on individual BCD images. The average ratio of
the flux density measured with 13\arcsec\ aperture photometry compared to
3-pixel (7.65\arcsec) PRF photometry is 1.371 with a dispersion of 0.011 for
108 sources brighter than 20~mJy. From the SSC web
pages\footnote{http://ssc.spitzer.caltech.edu/mips/apercorr}, the aperture
correction for a 13\arcsec\ aperture radius and the adopted sky annulus is a
1.167. The final flux densities were obtain by multiplying the PRF flux
densities by the product of these factors (1.600).

\begin{figure}
\includegraphics[angle=-90,scale=0.9]{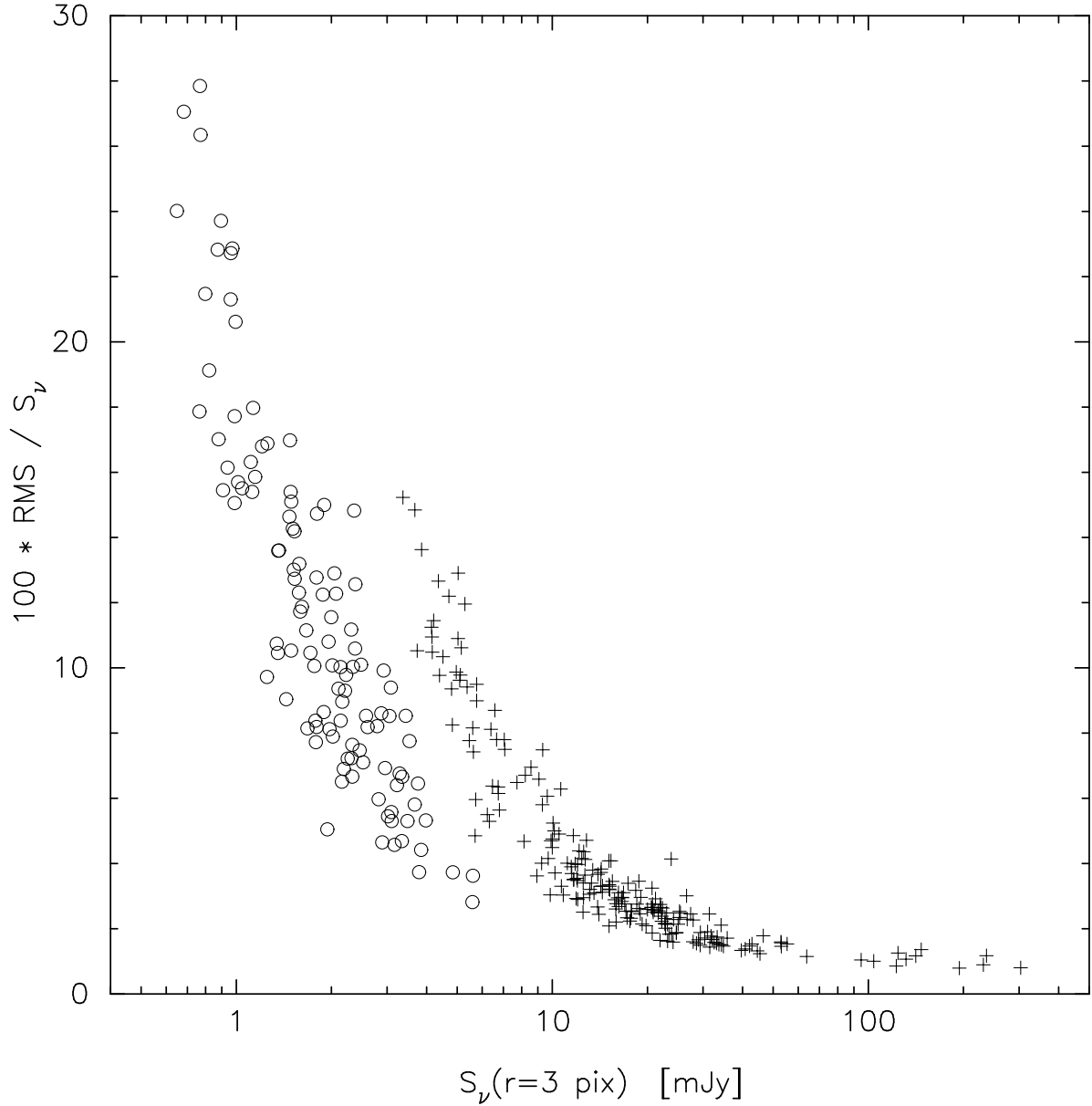}
\caption{
  \label{fig:rms_mips24}
  RMS repeatability of the MIPS~24\micron\ aperture photometry measured 
  in a 3~pixel radius on individual BCD images. Crosses represent sources
  observed with an exposure time of 3~sec, and open circles with 10~sec. We
  adopted a minimum uncertainty of 0.9\% based on the mean repeatability for
  stars brighter than 100~mJy.
}
\end{figure}

Internal uncertainties computed by APEX are often much smaller ($\ll 1\%$) 
than is assessed from repeated observations of the source. The minimum 
internal uncertainty was estimated based on photometric 
repeatability from aperture photometry. The normalized RMS of the 
MIPS~24\micron\ flux densities measured from {\it aperture} photometry 
on individual BCD images is presented in Figure~\ref{fig:rms_mips24}. 
For sources brighter than 100~mJy, the mean RMS repeatability is 0.9\% in
a 3 pixel aperture radius, which we adopted as the minimum uncertainty for the 
PRF photometric uncertainties.

The MIPS~24\micron\ photometry and internal uncertainties are presented in 
Table~\ref{tbl:phot}. The S13 images were processed with a calibration factor 
of 0.0447~MJy~sr$^{-1}$. Following \citet{Engelbracht07}, we adopt a
calibration uncertainty of 4\%.

\section{MIPS 70\micron}
\label{mips70}

We obtained MIPS~70\micron\ observations for 323 sources and extracted data 
for five stars (HD~17925, HD~72905, HD~202917, HD~216803, ScoPMS~214) 
from the \Spitzer\ archive. The FEPS observations were obtained in photometry
mode with an exposure time of 10~sec and the small field size dither pattern. A
single MIPS~70\micron\ image in this mode contains $32\times32$ pixels with a
scale of 9.8\arcsec~pixel$^{-1}$. The FEPS sources were centered on the left
half of the array which had the best sensitivity. The number of cycles per
source ranged between 2 and 10, where a cycle contains up to 12 dithered
images. The number of cycles were set based on the stellar distance and age to
reach the expected brightness of the outer Solar System dust level at that
stellar age \citep[see][]{Hillenbrand08}.

\subsection{Image Processing}
\label{mips70:images}

MIPS~70\micron\ images were processed with SSC pipeline version S13 that
removes the bias, subtracts a dark image, applies a flat field correction, and
linearizes the pixel response. Individual BCD images were mosaicked with
the Germanium Reprocessing Tools (GeRT) software package S14.0 version 1.1
developed at the SSC. The GeRT package performs column spatial filtering on 
the BCD images and then a time median filter to remove residual pixel response
variations. A $40''\times40''$ region centered on the source position, compared
to the PRF full-width-at-half-maximum (FWHM) size of 18\arcsec, was excluded 
when computing the column and time filtering such that the filtering process is
not biased by the presence of a bright source. Filtered images were formed
into mosaics with MOPEX \citep{Makovoz05}. Outlier pixels were rejected using
a 3$\sigma$ clipping threshold. 

\subsection{Photometry}
\label{mips70:phot}

Aperture photometry was performed on the MIPS~70\micron\ mosaics with a 
custom version of IDLPHOT. We adopted aperture photometry over PRF 
fitting photometry since most sources were not detected at 70\micron, and 
aperture photometry enables a straightforward interpretation of the
upper limits.

The adopted aperture radius of 16\arcsec\ (4 pixels on the coadded images), 
which corresponds to approximately
the FWHM size of the PRF, was chosen to optimize the signal to noise for faint
sources \citep[see, e.g.,][]{Naylor98}. The sky-level was computed as the mean
pixel value in a sky-annulus that extends from 40$''$ to 60$''$ after
performing the iterative clipping procedure described in \S\ref{irac:phot}. The
aperture was centered on the expected stellar position computed from the WCS
parameters contained in the FITS image headers, and no centroiding was
performed. Visual inspection of the 70\micron\ mosaics identified 19 images
where a point source was located within the outer sky annulus or the aperture
radius, but offset from the stellar position determined from 2MASS astrometry.
A PRF was fitted to the contaminating source and
subtracted from the image using MOPEX. These 19 sources are identified in
Table~\ref{tbl:phot}. Aperture photometry was recomputed on the PRF-subtracted
image.

The 70\micron\ photometric uncertainty was computed as
\begin{equation}
\sigma =
       (\eta_{\rm sky}\ \eta_{\rm corr})\
       (\Omega\ \Sigma_{\rm sky})\
       \sqrt{N_{\rm ap} + N_{\rm ap}^2/N_{\rm sky}}\label{eq:noise},
\end{equation}
where $\Sigma_{\rm sky}$ is the noise per pixel in units of surface brightness
as measured in the sky annulus, $\Omega$ is the solid angle of a pixel, $N_{\rm
ap}$ is the number of pixels in the aperture, $N_{\rm sky}$ is the number of
pixels in the sky annulus, and $\eta_{\rm sky}$ and $\eta_{\rm corr}$ correct
for correlated noise terms as described below. The total uncertainty is the
root-mean-square sum of two terms: the term proportional to $\sqrt{N_{\rm ap}}$
is the uncertainty from random fluctuations in the pixel noise summed over the
aperture, and the term proportional to $\sqrt{N_{\rm ap}^2/N_{\rm sky}}$
represents the uncertainty in the mean pixel noise from the sky annulus (often
assumed to be zero due to the large area over which one usually measures
the mean sky).

Two correction factors are needed to compute accurate uncertainties.
Because the 70\micron\ mosaics were sampled at a finer scale than the
raw images, the
noise between adjacent pixels is correlated. The factor $\eta_{\rm corr}$
accounts for the correlated noise, and was estimated as the ratio of the
pixel size in the raw images (9.8\arcsec) to that in the mosaics
(4\arcsec), or $\eta_{corr} = 2.5$.

The second correction factor, $\eta_{sky}$, accounts for 
systematic differences in the pixel noise between the aperture and 
sky annulus. Variations in the pixel noise as a function of position
across the mosaics were assessed by first scaling all 70\micron\
mosaics in the FEPS program to a common median value. The standard
deviation of each pixel in the stack of mosaic images was computed
after removing 35 images where the FEPS target was clearly detected.
The resulting image showed that the mosaic noise was higher along the
columns near the source position due to time-variable latent images
from the calibration stim flashes.  The pixel noise was estimated
to be 40\% higher in the aperture compared to the sky
annulus, and we adopt $\eta_{sky}$ = 1.40.

\begin{figure}
\includegraphics[angle=-90,scale=0.9]{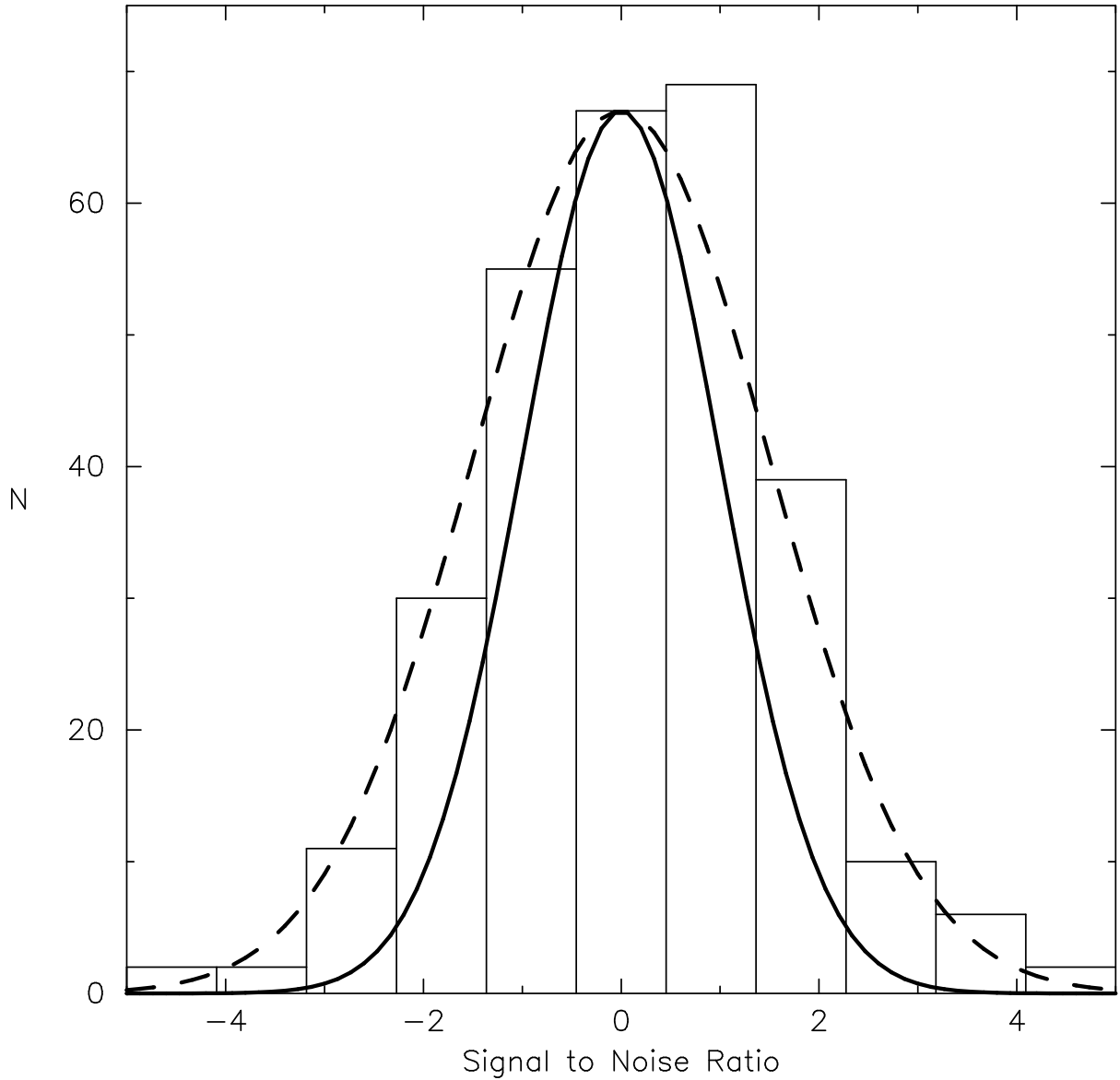}
\caption{
  \label{fig:mips70_snr}
  Histogram of the signal-to-noise ratio measured in a 16\arcsec\ radius
  aperture at the expected stellar position in the MIPS~70\micron\ mosaics.
  The solid curve shows the expected signal-to-noise distribution for gaussian
  noise (dispersion = 1.0) scaled to a peak value of N=67. The
  dashed curve shows a gaussian with a dispersion of 1.49. These results suggest
  that the 70\micron\ photometric uncertainties are underestimated by a factor
  of 1.49. The uncertainties in the 70\micron\ flux densities reported in
  Table~\ref{tbl:phot} have been scaled by this factor. 
}
\end{figure}

Figure~\ref{fig:mips70_snr} shows a histogram of the signal-to-noise ratio
(SNR) for the 70\micron\ photometry. Visual inspection of the mosaicked images
indicates that the majority of the FEPS sources have not been detected at
70\micron. The histogram in Figure~\ref{fig:mips70_snr} should then be a
gaussian with unit dispersion (solid curve in Fig.~\ref{fig:mips70_snr}) if
Equation~\ref{eq:noise} contains the dominant noise terms. In practice, the
observed SNR distribution is broader than the expected gaussian distribution
and includes SNR values as low as -4.2. 
As shown by the dotted curve in Figure~\ref{fig:mips70_snr}, a gaussian with a
dispersion of 1.49 adequately describes the observed distribution. The origin
of the apparent excess noise is unknown, but nonetheless, we have scaled the
photometric uncertainties for all sources by a factor of 1.49.

MIPS 70\micron\ photometry is calibrated to a theoretical PRF measured computed
over a $64'\times64'$ field \citep{Gordon07}. The aperture correction needed to
place the background-subtracted flux densities measured in a finite aperture on
the same scale as the theoretical PRF depends on the temperature of the
underlying source emission. In anticipation that the FEPS MIPS~70\micron\
observations did not detect the stellar photosphere in most cases and that
debris disks around solar-type stars have temperatures of \about 50-100~K,
aperture corrections were measured on a 100~K PRF \citep{Gordon07}. The 
aperture correction derived for our adopted aperture radius of 16\arcsec\ and 
sky annulus between 40 and 60$''$ is 1.766. By comparison, the SSC web
pages indicate that the
aperture of correction for a 3000~K and 15~K PRF is 1.741 and 1.884
respectively for the same 16\arcsec\ aperture radius and similar, but not
identical, background annulus of 39 to 65\arcsec.

The MIPS~70\micron\ photometry and internal uncertainties are presented in 
Table~\ref{tbl:phot}. The FEPS sources where the 70\micron\ photometry was
measured on PRF-subtracted images are marked in the table. The adopted
calibration factor is 702.0~MJy sr$^{-1}$ / (DN s$^{-1}$) with an uncertainty
of 7\% as reported on the SSC MIPS calibration web
pages\footnote{http://ssc.spitzer.caltech.edu/mips/calib}.

\section{IRS Low Resolution Spectra}
\label{irs}

Low resolution spectra ($\lambda/\delta\lambda$ \about 60--120) of the FEPS 
sources were obtained with IRS. Most sources were observed in the short-low 1
(SL1, 7.4-14.5\micron), long-low 2 (LL2, 14.0-21.3\micron), and long-low 1
(LL1, 19.5-38\micron) orders. A subset of sources were also observed in the
short-low 2 (SL2, 5.2-7.7\micron) order. The spectral coverage beyond
35$\mu$m suffers from low signal-to-noise and was discarded for all sources.
The source HD~191089 was observed by a GTO program (PID 2, P.I. J. Houck) and
was not included in the FEPS IRS observations. Also, HD~72905 and HD~216803
were observed in the SL2 order only by FEPS; the longer wavelengths for
HD~72905 were observed as part of a GTO program (PID 41, P.I. Rieke). 

Targets were acquired in the spectrograph slit using either high-accuracy 
IRS or PCRS peak-up with a 1$\sigma$ radial pointing uncertainty of 
0.4\arcsec\ and 0.14\arcsec\ respectively according to version 8.0 of the
\Spitzer\ Observing
Manual\footnote{http://ssc.spitzer.caltech.edu/documents/SOM}. 
The reconstructed pointing from the peakup observations differed from the 
requesting pointing by $>9$\arcsec\ for 5 sources: HD~80606, HD~139813,
HII~2881, HIP~42491, and RX~J1544.0-3311. We assumed that the
spectra for these 5 sources are not for the intended target. For HD~13974, the 
pointing offset was within the pointing accuracy of the IRS peakup, but the 
intensity of the SL1 spectrum is a 2.6$\times$ lower than expected by 
extrapolating the IRAC 8\micron\ photometry to 13\micron\ assuming a
$\nu^2$ spectrum. For R45, the extracted spectrum had a flux density less than
0~Jy for wavelengths $> 25$\micron. We have excluded the SL and LL spectra
for HD~13974 and R45, respectively.

Two nod positions per cycle were obtained for the IRS observations in standard
staring mode with a minimum of six cycles per target to allow rejection of bad
pixels and cosmic ray hits. Each spectral image comes with a bit-mask image
that marks potentially bad pixels. The data conditions identified by each bit
in the mask are described in the \Spitzer\ Data
Handbook\footnote{http://ssc.spitzer.caltech.edu/irs/dh}. Pixels marked with
bit 9 or higher were replaced with the average pixel value of an 8 pixel box
surrounding the bad pixel. This method for finding the mean pixel value
resembles Nagao-Matsuyama filtering \citep{Nagao79} and ensured edge
preservation in the source region of our spectral images. 

Source spectra were extracted from the droop intermediate data product from the
SSC pipeline version S13 for all but two sources. The spectra for MML~18 and 
ScoPMS~52 were reobserved since the initial observations had a failed peakup,
and the final spectra were extracted from S15 data products.
Background emission and stray-light were eliminated
by subtracting images of the two slit positions at which a target is observed
for each module and order. This resulted in a set of images containing a
positive and negative spectrum in each observed order. A straight-sided
(boxcar) aperture was used to extract the spectra for each nod-position and
cycle. 

We found that the source positioning has the expected 0.4\arcsec\ (1$\sigma$)
pointing accuracy, but that the targets are not positioned exactly on the
1/3-2/3 position along the slit.  The width of each aperture was determined by
two quantities: the maximum size of the PSF in each order, and the pointing
accuracy. The width of each aperture is chosen such that 99\% of the source
flux is within the aperture. To estimate the size of a point source we assume a
Gaussian PSF with a FWHM = 0.25 * lambda, where lambda is in microns and FWHM
in arcsec. Taking also the positioning constraints into account, the apertures
are widened an additional 2.4\arcsec\ (6$\sigma$), to ensure that the entire
source is always positioned within the aperture. The resulting extraction boxes
were 6 pixels (11.1\arcsec) and 5 pixels (25.4\arcsec) along the slit for the
short-low and long-low modules, respectively. 
Given that the slit width is only 2 pixels, pointing uncertainties in the
dispersion direction will dominate the error on the flux density.

Because spectra were extracted with custom apertures that differ from the
SSC processing, the spectral response function (SRF) had to be derived. We 
used a set of high signal-to-noise observations of bright calibration stars 
with model spectra provided by the SSC to derive the relative SRF, and then an 
internal calibration to determine the absolute flux calibration. Calibrating 
slit spectra suffers from uncertainties in the adopted
spectral model and flux losses due to pointing offsets of the slit
compared to the target. The FEPS Legacy program provides a unique opportunity 
to derive a good flux calibration for solar-type main-sequence stars since 
many stars do not exhibit emission from cool dust in the IRS wavelengths 
\citep{Carpenter08}. The SRFs
were determined for each order separately as the ratio of the observed spectrum
to a Kurucz model spectrum using calibration stars identified in the FEPS
program. The Kurucz model spectra were derived using the procedure outlined in
Appendix~\ref{kurucz}. Calibration stars were selected from the FEPS program by 
computing synthetic fluxes from the IRS spectra at wavelengths of 8, 13, 24,
and 33\micron\ and applying the following criteria:
(1) the flux density ratios of the synthetic photometry points at 8, 13, 24
and 33$\mu$m are within one sigma of the colors expected for stellar 
photospheres;
(2) there were no known peak-up problems during data acquisition;
(3) the spectra contains no artifacts from cosmic ray hits, hot or dead 
pixels; 
and 
(4) the spectra have among the best signal-to-noise for the specific order and 
ramp time to ensure high quality SRFs. The SRFs were derived from a set of 16
calibration stars for the SL1 and LL orders, and from a separate set of 10
stars for the SL2 order.

After the extraction of each spectrum and normalization by the SRF, a mean
spectrum over all slit positions and cycles was computed for each individual
order. The orders were then combined to form a single spectrum. In the regions
where the spectra of the individual orders overlap, the flux densities were
replaced by the mean flux density at each wavelength point. Internal
uncertainties per pixel were estimated as the standard deviation of the mean of
the repeated spectral observations. The SRF based on the bright calibration 
stars from the IRS instrument team (the spectra were extracted in an identical 
way to the FEPS sample) were then scaled to the SRF derived from the internal
calibration described above. This procedure ensures that the uncertainties
introduced by the adopted spectral model and flux losses due to pointing 
offsets of the slit are minimized and that the signal-to-noise ratio on the 
relative SRF is much better than that of the spectrum of any individual 
target.

The final calibrated spectra, excluding the problem spectra mentioned above,
are distributed in the electronic version of this article. Each data file
contains a header summarizing the observational parameters and four data
columns that list the wavelength in microns, the flux density and internal
uncertainty in Janskys, and the spectral order number.

\section{Source Confusion}
\label{confusion}

Infrared cirrus and extragalactic sources may contaminate the FEPS photometry
and create the appearance of an infrared excess. Since we anticipate that the
emission associated with the stellar photosphere or a circumstellar disk will
be nearly point-like and centered on the star at the typical distances in the
FEPS sample, potential contamination to the 24\micron\ or 70\micron\ photometry
can be identified from emission that is extended or offset from the stellar
position. 

We used the 2MASS catalog to represent the stellar position since most stars in
the FEPS sample do not exhibit an infrared excess in the $JHK_{\rm s}$ bands
\citep{Carpenter08}, and any such excess should be unresolved spatially. 
2MASS astrometry was corrected to the epoch of the \Spitzer\ observations based
on proper motions in the Tycho-2 \citep{Hog00} or UCAC2 \citep{Zacharias04}
astrometric catalogs. MIPS~24\micron\ source coordinates were computed from the
PRF centroid position and the WCS astrometric solution in the FITS
image headers. 

\begin{figure}
\includegraphics[angle=-90,scale=0.7]{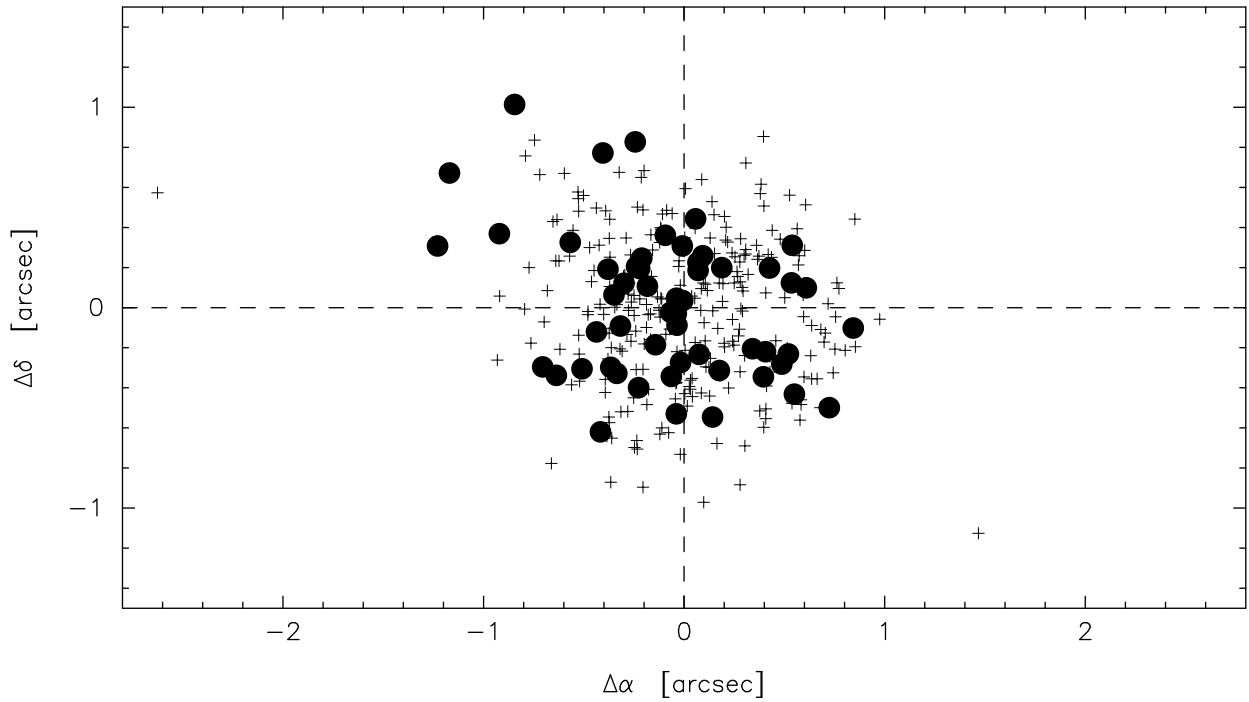}
\caption{
  \label{fig:mips24_coords}
  Angular offset between the 24\micron\ emission centroid 
  and the 2MASS position after correcting for proper motion and differences
  in epoch of observations. Filled circles represent FEPS stars that exhibit 
  an infrared excess in the IRS spectrum, and crosses represent sources 
  without a detectable IRS excess \citep{Carpenter08}.
}
\end{figure}

In Figure~\ref{fig:mips24_coords}, we show the angular
separation between the 2MASS and MIPS 24\micron\ astrometry, where solid
circles represent sources that exhibit an infrared excesses in the IRS spectra
and crosses indicate stars without an excess \citep[see][]{Carpenter08}. Two
sources have 24\micron\ positions that are offset by more than 1.8\arcsec\ from
the 2MASS coordinates, but neither exhibits an IRS infrared excess. Excluding
these two outliers, the dispersion in the right ascension and declination
offsets are 0.41\arcsec\ and 0.36\arcsec, respectively, with a radial
dispersion of 0.49\arcsec. The dispersion is
dominated by uncertainties in the \Spitzer\ astrometry since the typical
1$\sigma$ uncertainty in the 2MASS positions is \about 0.14\arcsec\
\citep{Skrutskie06}. Stars with infrared
excesses have a larger dispersion in the radial coordinate offsets than stars
without infrared excesses (0.30$''$ vs 0.23$''$), which can be attributed to 3
excess sources (HD~35850, HD~201219, and HD~209253) that have offsets of \about
$1.3$\arcsec. The 24\micron\ excess source with the largest angular offset,
which is HD~35850 at 1.35\arcsec, deviates from the 2MASS position by 
2.9$\sigma$ in right ascension and 1.9$\sigma$ in declination. We conclude
that for most FEPS sources, the 24\micron\ astrometry offsets relative to 
2MASS is similar for stars with and without an infrared excess. Potentially
three excesses sources may be contaminated by cirrus or extragalactic sources
to produce an unusually large offset (1.3\arcsec). However, we cannot rule out
pointing reconstruction errors since the two largest astrometric offsets 
are found around stars without infrared excesses.

\begin{figure}
\includegraphics[angle=-90,scale=0.9]{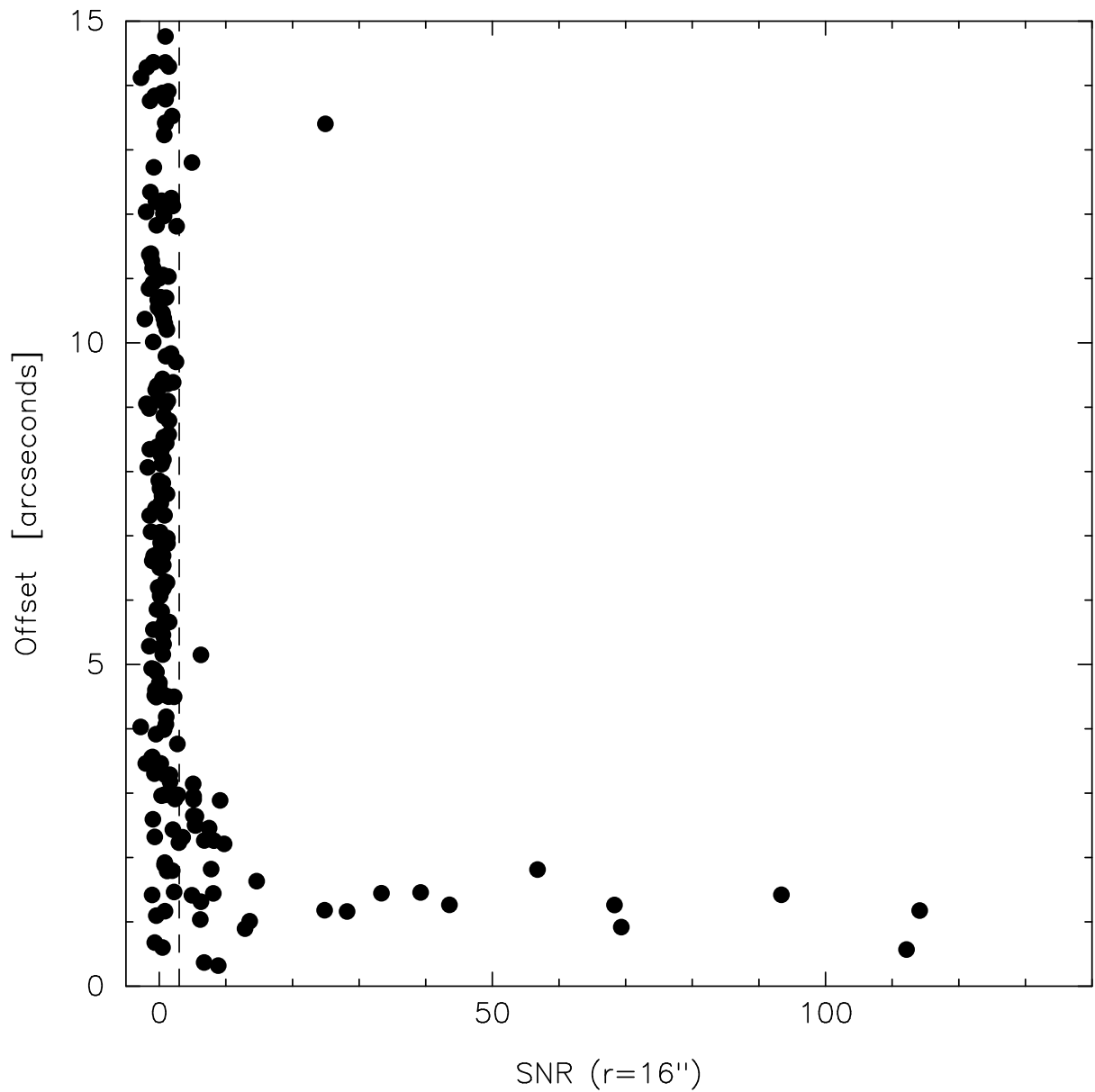}
\caption{
  \label{fig:mips70_coords}
  Angular offset between the 70\micron\ coordinates and the 2MASS stellar 
  position as a function of the 70\micron\ signal-to-noise ratio measured in a
  16\arcsec\ radius aperture. The 70\micron\ centroid was computed by fitting 
  a two-dimensional gaussian to a 44\arcsec$\times$44\arcsec\ region centered 
  on the
  stellar position. The vertical dashed line at SNR=3 indicates the minimum
  signal-to-noise ratio that defines a MIPS~70\micron\ detection. 2MASS 
  coordinates have been corrected to the \Spitzer\ epoch of observations 
  using published proper motions. 
} 
\end{figure}

The relative MIPS 70\micron\ and 2MASS astrometry was evaluated in a similar
manner. We computed the 70\micron\ emission centroid by fitting a
two-dimensional gaussian to a 44$\times$44\arcsec\ (11$\times$11 mosaicked
pixels) region centered on the expected stellar position. In
Figure~\ref{fig:mips70_coords}, we show the difference between the 70\micron\
and 2MASS astrometry as a function of the 70\micron\ SNR measured in a
16\arcsec\ aperture. For sources with SNR $\ge$ 3, the positional agreement is
better than 3.5\arcsec\ for all but three sources: HD~201219 (5.1\arcsec\
offset), HD~104467 (12.8\arcsec), and RX~J1111.7$-$7620 (13.4\arcsec).
RX~J1111.7$-$7620 is separated by 24.4\arcsec\ from the classical T~Tauri star
XX~Cha; these sources have comparable brightness at 70\micron\ and the gaussian
fit converged to a centroid intermediate between the two sources. The
70\micron\ detection toward HD~104467 is a point source clearly offset from the
stellar position. Given the large offset, we assume that the detected 
70\micron\ source is unrelated to the star. Finally, the HD~201219 70\micron\ 
mosaic contains two point sources separated by 20.9\arcsec\ that distorted the
gaussian fit. We fitted gaussians to both sources and determined that the
brighter of the two sources is 3.4\arcsec\ from the 2MASS position for
HD~201219, which is not unusual given the 70\micron\ SNR (5.9) for this source.
However, this source also exhibits one of the larger angular offsets between
the MIPS~24\micron\ astrometry and 2MASS.
While neither the 24\micron\ nor the 70\micron\ astrometry conclusively 
demonstrates that the MIPS photometry for HD~201219 is contaminated, it 
suggest that the photometry for this source should be used with caution.

\begin{figure}
\includegraphics[angle=-90,scale=0.9]{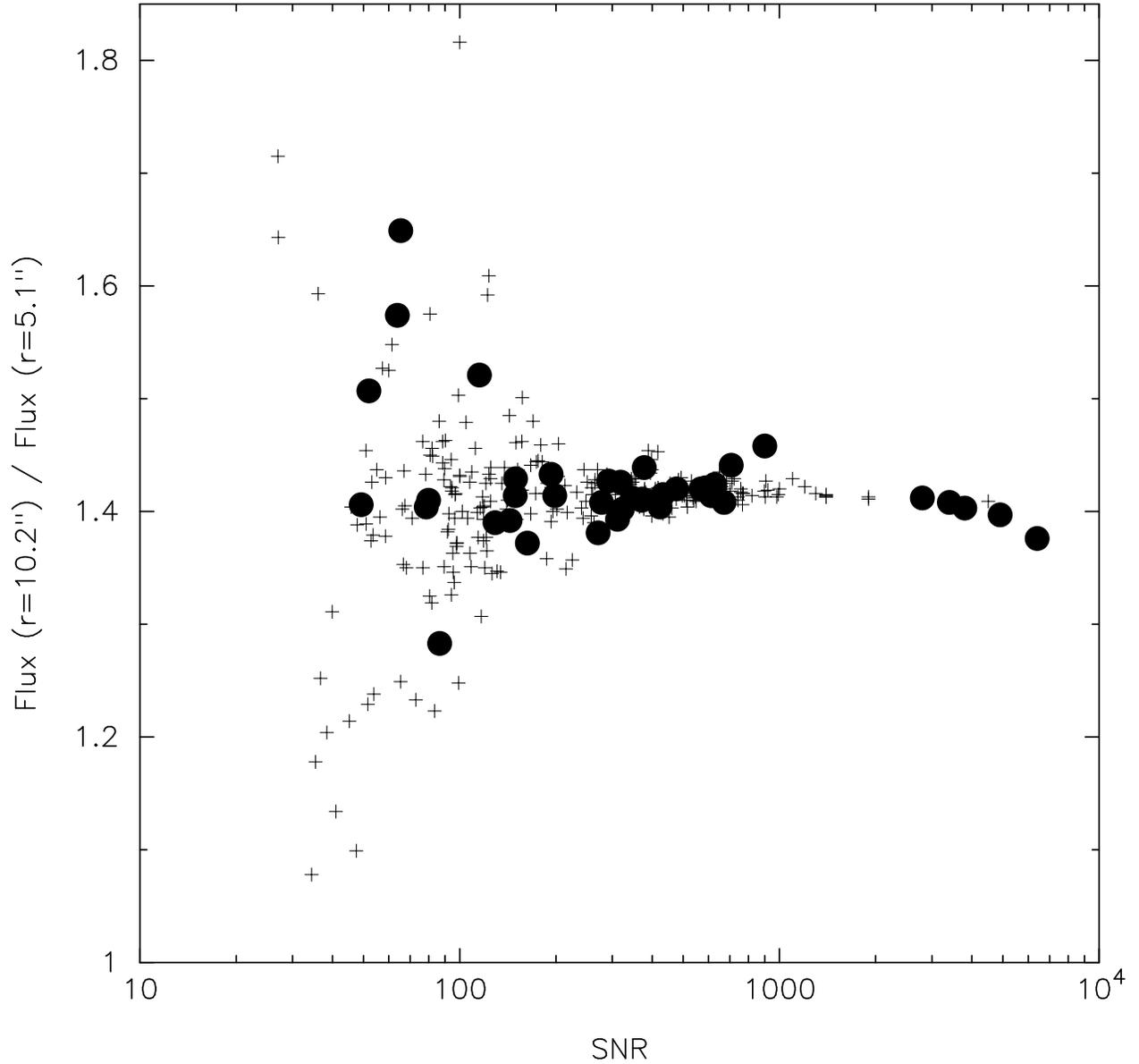}
\caption{
  \label{fig:mips24_cog}
  Ratio of the 24\micron\ flux density measured in a 10.2\arcsec\ radius
  aperture (= 4 pixels) to that in a 5.1\arcsec\ radius aperture (= 2 pixels)
  as a function of the signal-to-noise
  ratio of the 24\micron\ PRF photometry. Filled circles represent sources
  with a $\ge 3\sigma$ 24\micron\ excess confirmed by the IRS spectrum, and 
  crosses indicate sources without detectable 24\micron\ excesses
  \citep{Carpenter08}.
}
\end{figure}

To further search for possible contaminants in the MIPS photometry, we 
computed the
ratio of the flux measured in a large (10.2$''$ and 30$''$ for MIPS~24 and
70\micron, respectively) to a small (5.2$''$ and 16$''$ for MIPS 24 and
70\micron) aperture radius. A contaminating object or extended emission will
create an anomalous ratio between aperture sizes. In
Figure~\ref{fig:mips24_cog}, we show the flux ratio measured in a large
aperture to that in a small aperture as a function of the signal to noise ratio
for the MIPS 24\micron\ photometry. The scatter in the flux ratio is similar
for sources with (solid circles) and without (crosses) 24\micron\ excesses. For
SNR $>$ 300, the source with the most discrepant flux ratio at 24\micron\
relative to the other sources is HD~107146 at SNR=900. Several studies have
demonstrated that this source is surrounded by a circumstellar disk
\citep{Ardila05,Williams04,Carpenter05} and the observed flux ratio suggests
that the source may be extended at 24\micron. Sources with a 24\micron\ SNR
ratio less than 100 exhibit a larger scatter in flux ratios. The range
of values is similar for sources with and without infrared excesses, and
suggests that the scatter can be attributed to lower signal-to-noise in the
larger photometric aperture.

\begin{figure}
\includegraphics[angle=-90,scale=0.9]{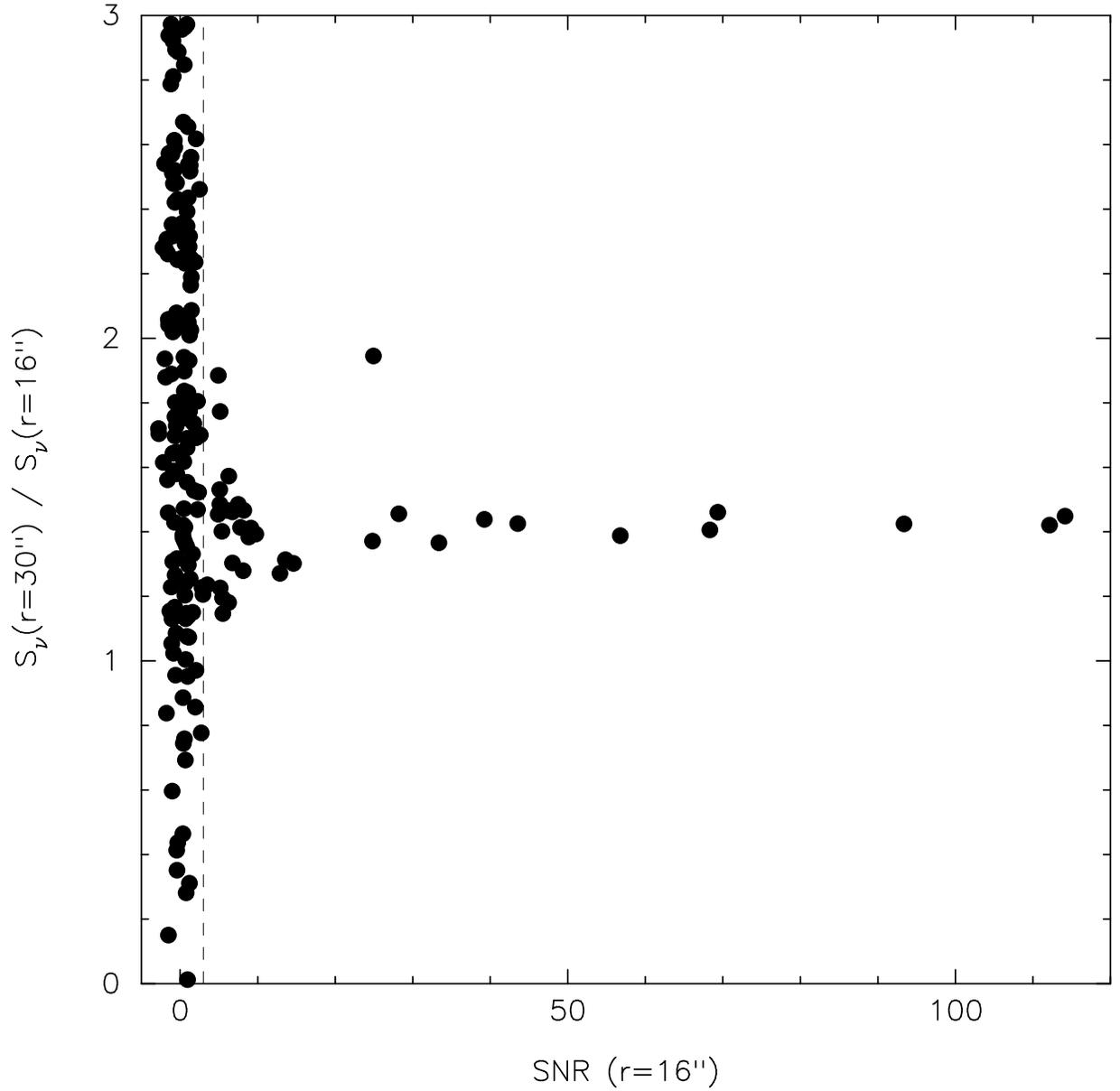}
\caption{
  \label{fig:mips70_cog}
  Ratio of the 70\micron\ flux density measured in a 30\arcsec\ radius aperture
  to that in a 16\arcsec\ radius aperture as a function of the signal to
  noise ratio. The vertical dashed line is drawn at SNR=3. The two sources
  with SNR $>3$ and flux density ratios greater than 1.8 have a nearby
  source that partially overlap the source aperture. These two contaminating
  sources were PRF-subtracted before performing the final photometry.
}
\end{figure}

In Figure~\ref{fig:mips70_cog}, we show the flux ratio in the two aperture
sizes as a function of the signal to noise ratio for the MIPS 70\micron\
photometry. Two sources (HD~104467 and RX~J1111.7$-$7620) with SNR $>$ 3 have
anomalously large ratios ($>$ 1.8). As discussed above, the initial photometry 
for these sources were contaminated by a nearby object, and the nearby source 
was PRF-subtracted before performing the final photometry. A third source
(HD~216803) has a flux ratio just under 1.8. The 70\micron\ emission for this
object is centered on the stellar position to within 3\arcsec, and the observed
70\micron\ emission is consistent with the expected stellar photosphere.

In summary, we conclude that the astrometry and curve-of-growth for most
sources are consistent with point source emission centered on the stellar
position. No compelling evidence exists that contaminants systematically
influence the 24 photometry. At 70\micron, contaminants needed to be removed
for a few sources before measuring the final photometry, and the sources are
noted in Table~\ref{tbl:phot}. These results do not exclude the possibility
that the photometry for some sources may be contaminated, but any such
contamination must be present in a minority of sources. 

\section{Cross-Instrument Calibration}
\label{crosscal}

In this section, we use the FEPS data to examine the cross-instrument 
calibration. We first analyze the 24\micron\ to 8\micron\ flux 
density ratio, which plays a prominent role in a companion paper to identify 
sources that exhibit excess emission from circumstellar dust 
\citep{Carpenter08}. We then compare the IRS and MIPS~24\micron\ calibration.

\subsection{IRAC 8\micron\ vs. MIPS~24\micron}

The observed IRAC 8\micron\ flux density is consistent with photospheric
emission for most stars in the FEPS sample \citep{Carpenter08}. The 24\micron\
to 8\micron\ flux ratio then is diagnostic of sources that exhibit
circumstellar dust emission at 24\micron. The precision to which this ratio can
identify excesses depends on the relative calibration stability of the IRAC 
and MIPS instruments over time and between the various observing modes. 

The primary difference between observations of different stars is the exposure 
time for individual IRAC and MIPS images. We first examine 
the relative stability of the MIPS 24\micron\ calibration by selecting stars 
in the FEPS program that were observed with the same IRAC frame-time, but 
different MIPS exposure times. We selected 48 stars in the FEPS sample that 
a) have been observed with IRAC frame-times of 0.10~sec,
b) do not show evidence for more than a 2$\sigma$ infrared excesses in 
   the FEPS IRS spectra \citep{Carpenter08} to ensure the 24\micron\ emission 
   is from the photosphere,
c) the variation in the encircled energy with aperture radius in the IRAC
   images is consistent with a point source (see \S\ref{irac:phot}),
and
d) the dereddened $J-K_{\rm s}$ color is less than 0.7~mag to remove the
   intrinsically reddest stars in the FEPS sample.
We used a 0.10~sec IRAC frame-time to obtain the largest sample of stars
observed with different MIPS exposure times. 

\begin{figure}
\includegraphics[angle=0,scale=0.8]{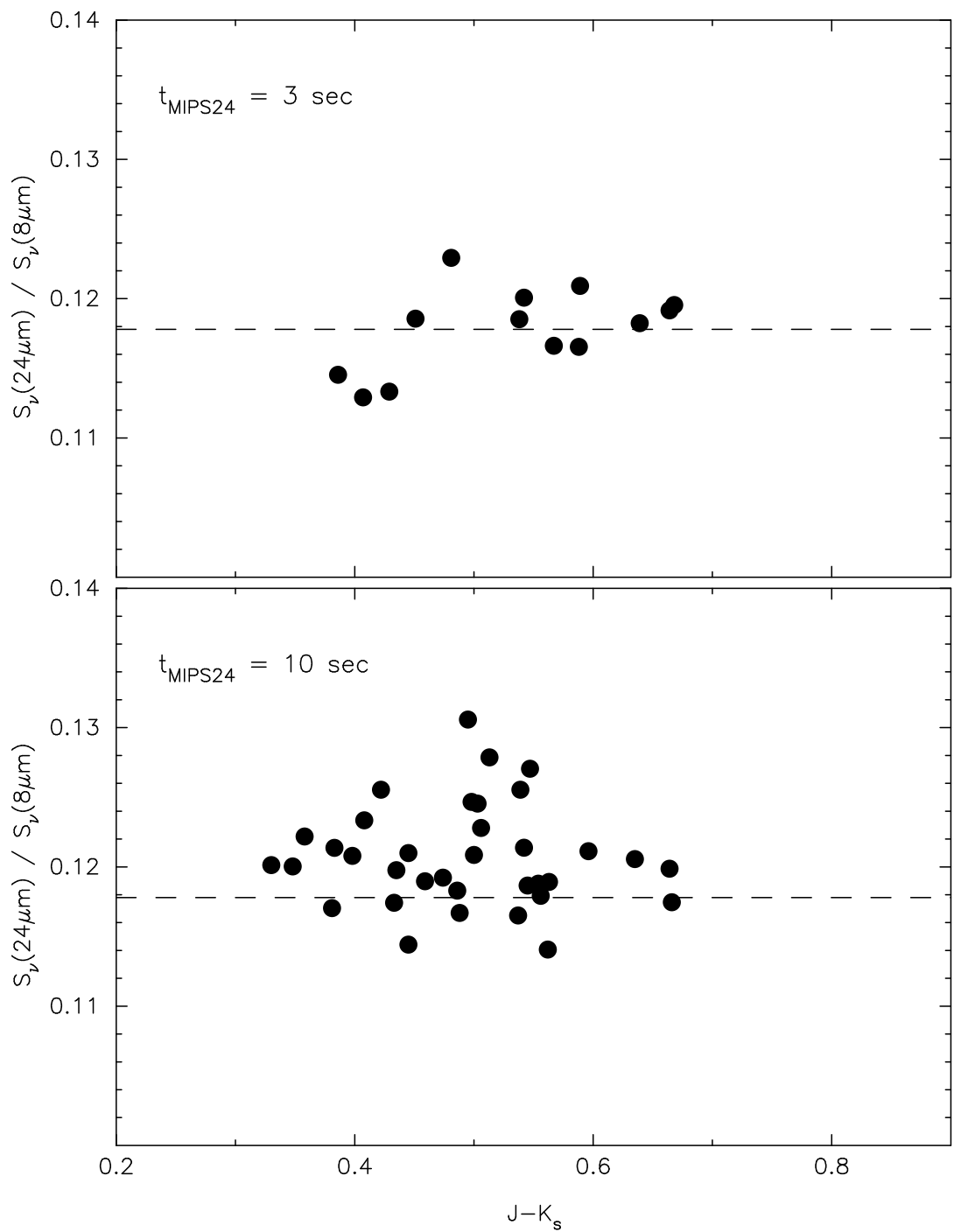}
\caption{
  \label{fig:mips24_offset}
  Ratio of 24\micron\ to 8\micron\ flux densities (\RM) plotted as a function 
  of the $J-K_{\rm s}$ color for stars observed with a 0.10~sec IRAC frame-time
  that do not have a IRS excess (see text for a complete description of the 
  selection criteria). The top panel shows the results for stars observed with
  a MIPS~24\micron\ exposure time of 3~sec, and the bottom panel for 10~sec 
  exposure time. The dashed line shows the mean flux ratios for the 
  3~sec MIPS data. The ratio of the mean value of \RM\ in the 3~sec MIPS data 
  to the 10~sec data is $0.976 \pm 0.008$.
}
\end{figure}

In Figure~\ref{fig:mips24_offset}, we plot the 24\micron\ to 8\micron\ flux
density ratio ($\equiv$ \RM) for MIPS 3~sec (top panel) and 10~sec (bottom
panel) exposure times versus the dereddened 2MASS $J-K_{\rm s}$ color using the
extinction estimates derived in \S\ref{stellar:av}. The two sample of stars 
span similar ranges of dereddened $J-K_{\rm s}$ colors, and 
we assume that the two samples also share the same intrinsic photospheric
$[8]-[24]$ color. The Shapiro-Wilk statistic ($\equiv p_{\rm SW}$) indicates 
that the distribution of observed data points about the mean is consistent with
a normal distribution for each sample ($p_{\rm SW}=0.76$ for the 3-sec MIPS
data and $p_{\rm SW}=0.31$ for the 10-sec data). The Student's $t$-test
then can be used to compare the mean values of \RM\ for the 3- and 10-sec MIPS
data. The probability from the $t$-test that the mean values of \RM\ for the
two samples are consistent with each other is 0.009. The ratio of the mean
value of \RM\ in a 3~sec MIPS exposure to that in a 10~sec exposure is
$0.976\pm0.008$, where the uncertainty was computed as the standard deviation
of the mean. These results suggest that the mean \RM\ value is higher for the
10-sec MIPS data on average compared to the 3-sec data.

\citet{Engelbracht07} measured directly any MIPS~24\micron\ calibration offsets
by observing a sample of 11 stars with 3, 10, and 30~sec MIPS exposure times.
They also found that the measured flux densities were larger on average in
10~sec exposure data compared to 3~sec observations. However, the magnitude of 
their offset (1\%) is 2.4 times smaller than the offset derived from the FEPS 
data. 
While the reduction procedure adopted here attempted to follow that recommended
by \citet{Engelbracht07}, our data processing was nonetheless performed using
SSC products and custom software that could account for the different
results. Also, we adopted PRF-fitting photometry, while \citet{Engelbracht07}
used aperture photometry. As a check of our data reduction methods, we 
compared our photometry with the results from \citet{Rieke08}, who used the
pipeline described in \citet{Engelbracht07} to process data for 31 FEPS sources
that were observed with 3~sec exposure times. For these 31 stars, the median
difference between the flux densities measured by FEPS and
\citet{Rieke08} is 0.0\% with a dispersion of 2.6\%. Therefore our data 
reduction procedures for at least the 3~sec exposure data yields photometry
consistent with the \citet{Engelbracht07} processing, but no 
independent check is available for the 10~sec MIPS data.

\begin{figure}
\includegraphics[angle=0,scale=0.8]{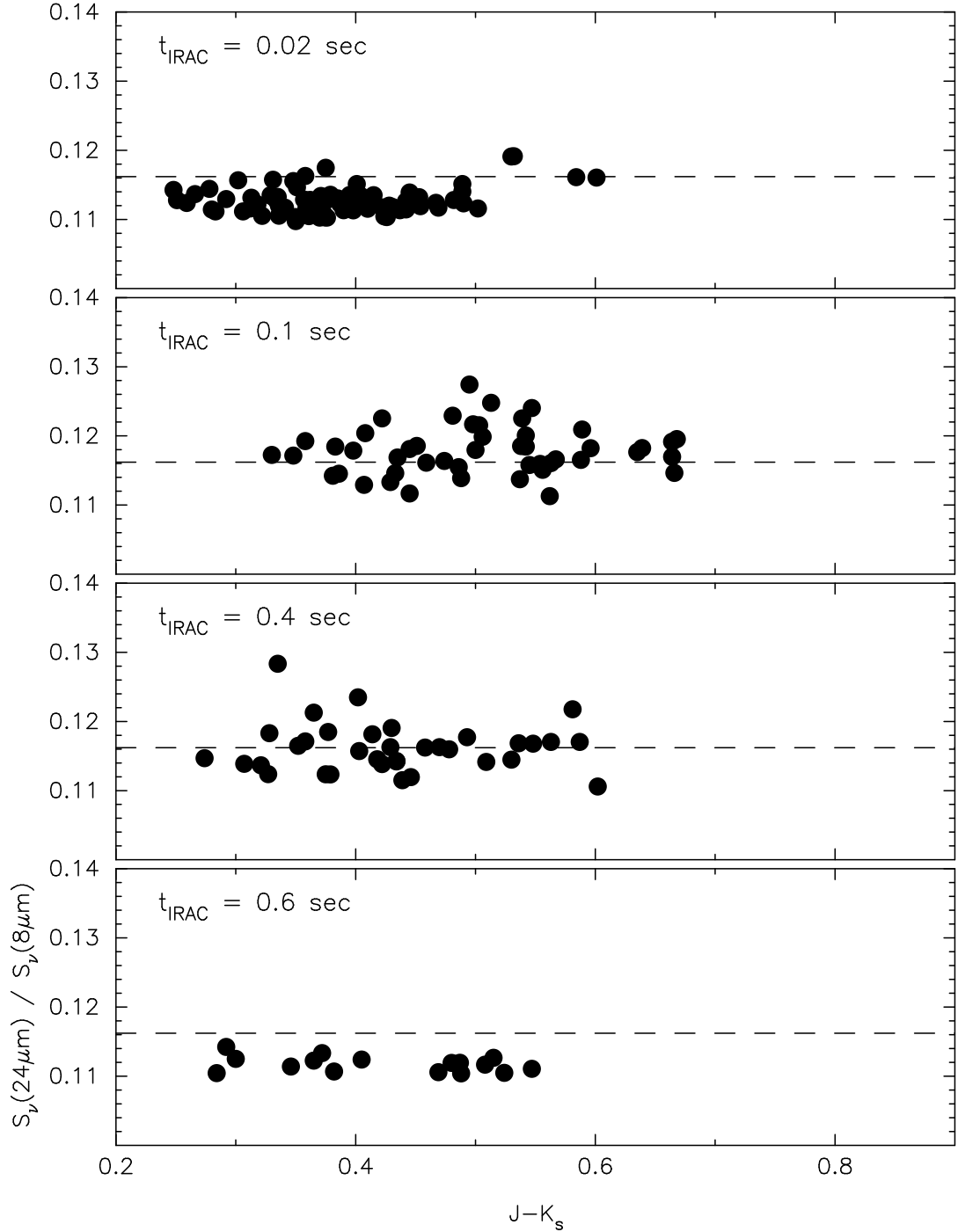}
\caption{
  \label{fig:irac4_offset}
  Ratio of 24\micron\ to 8\micron\ flux densities plotted versus dereddened
  $J-K_{\rm s}$ color for IRAC frame-times of 0.02, 0.1, 0.4, and 0.6~sec. 
  The dashed line in each panel shows the mean flux ratio for the 
  0.4~sec IRAC data. These results suggest that the observed
  24\micron\ to 8\micron\ flux ratio varies with IRAC frame-time.
}
\end{figure}

\begin{figure}
\includegraphics[angle=0,scale=0.8]{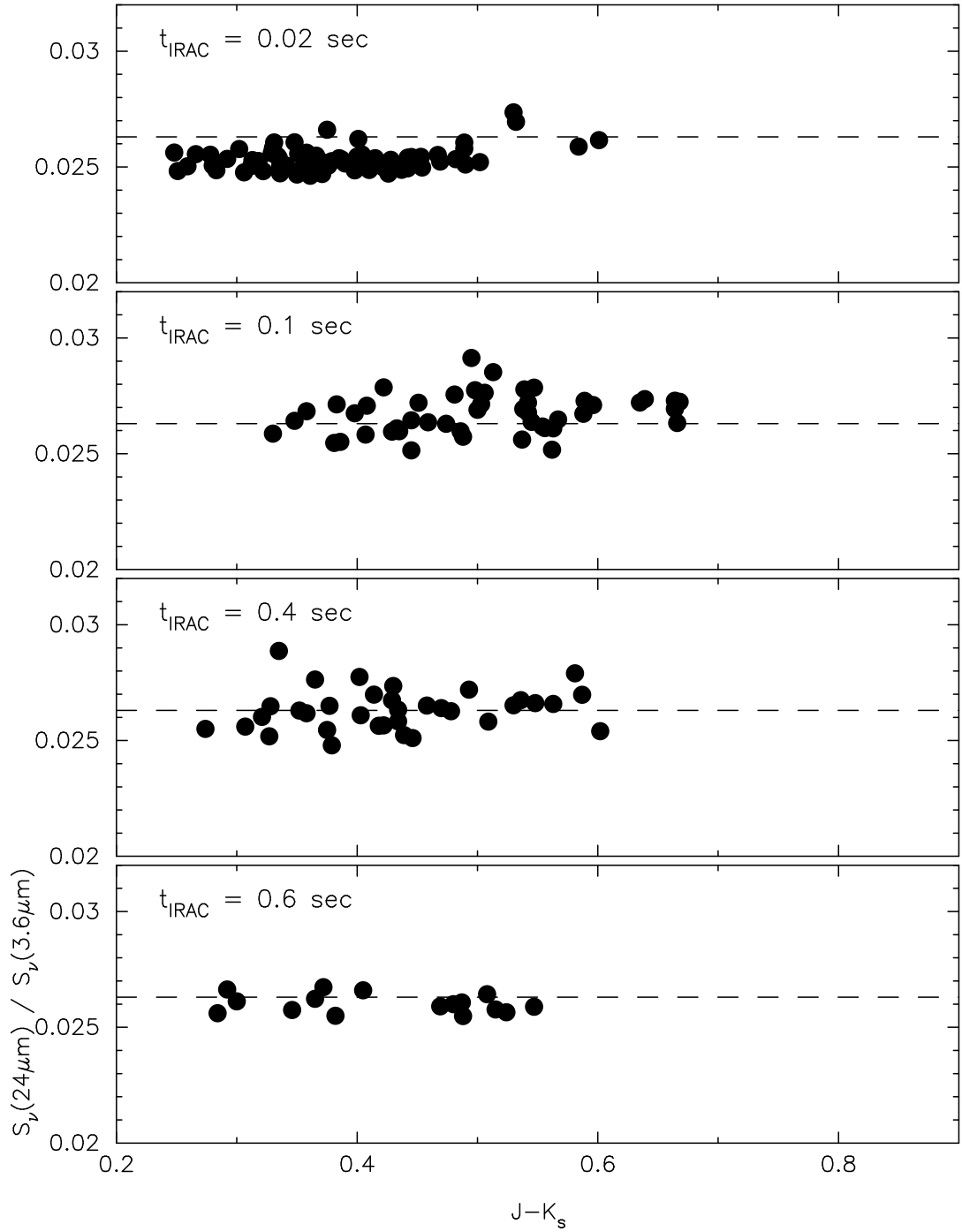}
\caption{
  \label{fig:irac1_offset}
  Same as Fig.~\ref{fig:irac4_offset}, but for the IRAC 3.6\micron\ band.
}
\end{figure}

\begin{figure}
\includegraphics[angle=0,scale=0.8]{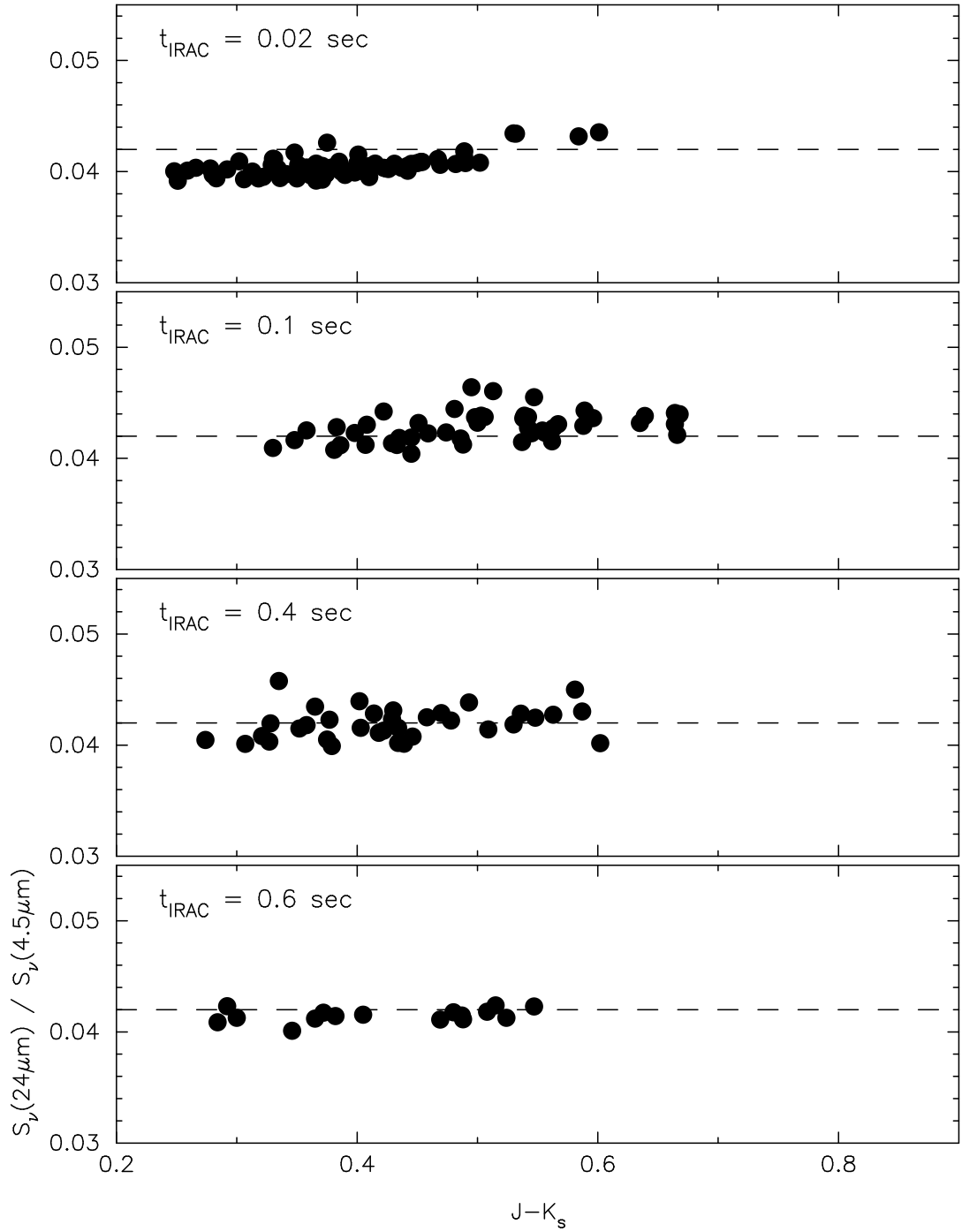}
\caption{
  \label{fig:irac2_offset}
  Same as Fig.~\ref{fig:irac4_offset}, but for the IRAC 4.5\micron\ band.
}
\end{figure}

We now consider the relative flux calibration for stars with different IRAC
exposure times. In Figure~\ref{fig:irac4_offset}, we plot the 24\micron\ to
8\micron\ flux density ratio versus dereddened $J-K_{\rm s}$ color for stars
observed with various IRAC frame-times. The MIPS 24\micron\ photometry obtained
with 10~sec exposure times have been scaled by a factor of 0.976 based upon the
analysis above since the MIPS calibration is tied mainly to data obtained with
3~sec exposure times \citep{Engelbracht07}. As shown in the figure, systematic
differences are present in the mean flux density ratio between the various IRAC
frame-times. Offsets are present even if the 10~sec MIPS 24\micron\ data
are not scaled, but the magnitude of the offset changes. We adopt the 0.4~sec
frame-time as the fiducial calibration since the calibration of the 0.4~sec
sub-array data and the full-array data are the same to within 1\% (see 
\S\ref{irac:phot}). A multiplicative scale factor of $0.971\pm0.005$ must be
applied to the 0.02~sec IRAC frame-time data to force agreement with the
0.4~sec data, $1.014\pm0.007$ for the 0.1~sec data, and $0.962\pm0.006$ for the
0.6~sec data, where the uncertainties are the standard deviation of the mean.
In Figures~\ref{fig:irac1_offset} and \ref{fig:irac2_offset}, we present a
similar analysis for the IRAC 3.6 and 4.5\micron\ bands which demonstrates that
offsets are also present in these bands. Since only 5 FEPS stars were observed
in the IRAC 5.8\micron\ band, we were unable to derive offsets for that band. 

Table~\ref{tbl:irac_offsets} summarizes the multiplicative factors that must 
be applied to the flux ratios as a function of frame-time to scale the 
calibration to the 0.4~sec frame-time data. The offsets are similar in the 3
bands for a given frame-time, although the offset in the 0.6~sec frame-time
data may be larger for IRAC~8\micron\ than in the 3.6 and 4.5\micron\ bands. We
consider these correction factors preliminary since they have not yet been
verified by observing the same star with different frame-times. No
corrections for any integration-dependent calibrations have been applied to the
photometry in Table~\ref{tbl:phot}, but the frame-times are listed to enable
the corrections to be applied by the reader.

\subsection{IRS vs. MIPS~24\micron}

The IRS spectral coverage encompasses the spectral response of the 
MIPS~24\micron\ bandpass. To compare the relative calibration of
the two instruments, we computed synthetic 24\micron\ photometry from the
IRS spectrum and the MIPS~24\micron\ spectral response using the procedure
described in Appendix~\ref{kurucz:synthetic}. 

In Figure~\ref{fig:irs_mips24}, we plot the percent difference between the IRS
synthetic photometry and MIPS~24\micron\ photometry as a function of the
MIPS~24\micron\ flux density. No exposure-time dependent corrections have been
applied to the MIPS~24\micron\ flux densities for this analysis. For sources
brighter than 10~mJy, which have the highest signal to noise, the median
difference in the 24\micron\ flux densities between the IRS spectra and the
MIPS photometry is 2.1\%. The median difference for sources between 3 and
10~mJy is $-1.6$\%. These differences are within the 1$\sigma$ calibration
uncertainty for both MIPS \citep[4\%; ][]{Engelbracht07} and IRS ($>5$\%;
Infrared Spectrograph Data Handbook Version 3.1). However, for individual
sources, the difference between the MIPS and IRS flux densities are larger than
expected based on the quantifiable internal uncertainties. One significant
discrepancy is ScoPMS~52, where the IRS 24\micron\ flux density is 63\% higher
than the MIPS~24\micron\ flux density. Inspection of the MIPS 24\micron\ image
shows that there is a source 18\arcsec\ away that is an order of magnitude
brighter than ScoPMS~52 \citep[see][]{Bouwman08}, and this source likely
contributes flux to the IRS spectrum.

\section{Summary}
\label{summary}

The FEPS \Spitzer\ Legacy program was designed to obtain infrared photometry
from 3.6 to 160\micron\ and low resolution spectra from 5 to 35\micron\
for 328 solar-type stars spanning ages 
from 3~Myr to 3~Gyr. The broad goal of FEPS was to determine the 
incidence of circumstellar disks and place the results in context with the 
expected evolution of our Solar System. An essential component of this study
was to construct carefully calibrated spectral energy distributions.
Here, we outline the data reduction procedures adopted by the 
FEPS team to obtain accurate and well-characterized \Spitzer\ photometry and 
spectra.

The adopted image processing steps for the IRAC, MIPS, and IRS data closely 
follow the recommended procedures by the \Spitzer\ Science Center and 
\Spitzer\ 
Instrument Teams. We describe in detail the data reduction methods for each
instrument and the procedures used to validate the data products. We present
in Table~\ref{tbl:phot} the measured IRAC (3.6, 4.5, and 8\micron\ bands) and 
MIPS (24 and 70\micron) flux densities and uncertainties. The extracted, 
calibrated IRS spectra are available electronically. 

\acknowledgements

JMC thanks Dave Frayer, Sean Carey, Bill Reach, Jason Surace, and 
the staff at the \Spitzer\ Science Center for patiently answering numerous 
questions regarding \Spitzer\ data. We are grateful to the anonymous referee 
and George Rieke for providing valuable comments. We also thank 
Debbie Padgett, Tim Brooke, Dan Watson, Pat Morris, and the rest of the FEPS
team for their many contributions throughout this project. This work is 
based on observations made with the {\it Spitzer} Space
Telescope, which is operated by JPL/Caltech under a contract with NASA. The
program made use of data and resources from the FEPS project, which receives
support from NASA contracts 1224768, 1224634, and 1224566 administered through
JPL. This research made use of the SIMBAD database, operated at CDS,
Strasbourg, France, as well as data products from the Two Micron
All Sky Survey, which is a joint project of the U. Massachusetts and the
Infrared Processing and Analysis Center/Caltech, funded by NASA and the NSF.

{}

\appendix

\section{Stellar Photometry}
\label{newphot}

The FEPS team obtained optical photometry in the $BVRI$ broad-band filters 
for 45 stars. Observations were obtained with the 61$''$ Kuiper Telescope on
2003 May 8 and 2003 September 29-30, and the CTIO 0.9~m telescope on 2004
March 18-21.

The Kuiper observations used a 2048$\times$2048 pixel CCD with a pixel scale of
0.45$''$~pixel$^{-1}$. Images were processed by subtracting the bias, dividing
by a ``master'' flat field created from sky observations to remove large scale
response variations over the CCD, and dividing by a dome flat to remove
pixel-to-pixel variations. The CTIO observations were performed with a
2048$\times$2048 CCD and a pixel scale of 0.40$''$~pixel$^{-1}$. The CCD is
read out with different amplifiers for each quadrant. Each quadrant was bias
subtracted and divided by dome flats.

Photometry was measured using aperture photometry with a sky annulus that
extended from 20 to 30 pixels, and an aperture radius of 11 and 12 pixels for 
the CTIO and Kuiper images, respectively. Eight FEPS stars had a nearby source 
in projection, and a smaller aperture radius between 2-3 pixels to isolate the
photometry to the FEPS target. The observations were calibrated by observing
multiple standard stars from \citet{Landolt92} to solve for the airmass 
coefficient, the photometric zero point, and color terms to place the 
photometry on the Johnson-Cousins photometric system. Total photometric 
uncertainties were computed as the root-mean-square sum of internal 
photometric uncertainties, the zero point, and the color terms.
The photometry for the 45 sources are presented in Table~\ref{tbl:newphot}.

\section{Stellar Properties}
\label{stellar}

In this section, we describe the procedure to assign estimates of the
visual extinction, surface gravity, metallicity, and effective temperature
for each star in the FEPS sample. These derived parameters were used in 
several FEPS studies, and served as initial estimates for the Kurucz model 
fitting (see Appendix~\ref{kurucz}).

\subsection{Visual Extinction}
\label{stellar:av}

Distances to the FEPS targets extend upwards of 343~pc and the extinction from
the interstellar medium may be non-negligible. The visual extinction toward 
individual stars was estimated from one of the following techniques in
priority order:
1) proximity within the Local Bubble,
2) as a member of stellar cluster that has been extensively studied previously;
3) color excess at optical and near-infrared wavelengths;
and
4) a galactic extinction model.
We now describe each of these techniques.

Stars within the Local Bubble are expected to have small extinction at
visual wavelengths. The size of the Local Bubble has been measured by observing
interstellar absorption lines toward stars with known distances, and then
determining the column density as a function of distance. \citet{Welsh98}
present an analysis of Na~I column density measurements toward stars with
Hipparcos distance estimates, and they found that the visual extinction is less
than 0.01~mag out to a distance of $d=75$~pc. We adopted an extinction of 0~mag
for the 169 stars in the FEPS sample where $d + 3\Delta d \le 75$~pc,
where $\Delta d$ is the $1\sigma$ distance uncertainty
\citep[see][for a discussion on the distance determinations]{Meyer06}.

The visual extinction toward the clusters in the FEPS sample has been
extensively studied in the literature. For the Hyades, \citet{Taylor06} place 
an upper limit at 95\% confidence of $E(B-V) = 0.001$~mag, and we 
adopt $\rm{A_V} = 0$~mag. \citet{Breger86} compiled spectral types and optical 
photometry for about 120 Pleiades members and derived $E(B-V)$ = 0.04 on 
average, but with lower reddening to the east of the cluster (0.03~mag) 
compared to the west (0.06~mag). Assuming a factor of 3.1 to convert the 
$B-V$ reddening to visual extinction, we adopt a constant value of 0.12 mag 
for the Pleiades stars. Following \citet{Pinsonneault88}, we adopt an
average $E(B-V) = 0.10$ \citep[see][]{Crawford74,Prosser94}, or 
$\rm{A_V} = 0.31$~mag, for Alpha~Per. For IC~2602, we adopt a visual
extinction of 0.12~mag \citep{Whiteoak61}.

Many of the FEPS stars are field objects that have distances greater than 
75~pc. The visual extinction for these stars was computed from the color excess
given the published spectral types \citep[see][]{Meyer06} and 
observed colors. Optical (Johnson $B$ and $V$, Tycho $B_{\rm T}$ and 
$V_{\rm T}$) and near-infrared (2MASS $J$, $H$, and $K_s$) photometry were 
compiled from the literature (see Appendix~\ref{kurucz:phot}) or measured by 
the FEPS team (see Appendix~\ref{newphot}). 
The intrinsic colors as a function of spectral type were compiled from the 
literature by cross-correlating the Hipparcos catalog with the Michigan 
Spectral Catalog, Tycho-2, and 2MASS. The positional match between the
Tycho-2 and Michigan spectral atlas from \citet{Wright03} was used as a 
starting point. Only 2MASS sources with a PH\_QUAL flag of AAA and a 
confusion flag of 000 were used. A photometric uncertainty of less
$\le$ 0.072~mag (i.e. signal-to-noise ratio $> 15$) was required in each 
photometric band. The average color was then 
computed as a function of spectral type for stars within 75~pc for $B$ and
$V$ photometry, and within 100~pc for colors involving $J$, $H$, and $K_s$.
In computing the average colors, individual measurements were weighted by the
inverse variance of the measurements, and outliers from poor photometry or 
spectral types were removed in an iterative sigma-clipping procedure.
Table~\ref{tbl:colors} lists the adopted intrinsic colors for the relevant
spectral types in the FEPS sample, the dispersion in the observed colors,
and the number of stars that met the above criteria.
Color excesses were computed from the observed $(B-V)_{\rm Johnson}$,
$(B-V)_{\rm Tycho}$, $V_{\rm Tycho}-K_{\rm s}$, and $J-K_{\rm s}$ colors
and the intrinsic colors listed in Table~\ref{tbl:colors}. Intrinsic
$(B-V)_{\rm Johnson}$ colors were computed from the $(B-V)_{\rm Tycho}$
colors and the Tycho-to-Johnson transformation equations in
\citet{Mamajek02,Mamajek06}. The visual extinction was estimated for each
color using the extinction law compiled by \citet{Mathis90}, and the
weighted mean was adopted as the extinction.

For 9 stars, the visual extinction could not be estimated with the above
techniques since either a spectral type was not available, or the computed
extinction was unphysical (i.e. $\rm{A_V} < 0$~mag). In the latter case, it is
presumed that the photometry was poor or the spectral type is erroneous.
For these stars, we estimated the extinction using the \citet{Sandage72} 
extinction model assuming an exponential disk \citep[see][]{Chen98}.
The adopted extinction values are listed in Table~\ref{tbl:prop}.

\subsection{Stellar Effective Temperature, Surface Gravity, and Metallicity}
\label{stellar:logg}

The stellar effective temperature, surface gravity, and metallicity are needed 
to fit the Kurucz model atmospheres (see Appendix~\ref{kurucz}). This section 
summarizes the procedure to estimate these properties for the FEPS sample. 
The procedure depends on the stellar age, as solar-mass stars younger than 
\about 100~Myr are contracting toward the main sequence and the surface 
gravity varies with age.

Stars older than 100~Myr in the FEPS sample were considered to be main-sequence
stars and were assigned a surface gravity of ${\rm log}~g = 4.50$~g~cm$^{-2}$.
Stellar effective temperatures were estimated from the $B-V$ and $V-K$ 
versus temperature relations derived by \citet{Houdashelt00} after
dereddening the observed photometry (see \S\ref{stellar:av}). If the 
temperature uncertainty derived from the 
photometry is larger than 130~K, the temperature was instead computed from a
temperature vs. spectral type relation using the colors listed in 
Table~\ref{tbl:colors}, the Tycho-to-Johnson color transformations from
\citet{Mamajek02,Mamajek06}, and the \citet{Houdashelt00} color-temperature
relations. A limit of 130~K was adopted since that is approximately the
temperature uncertainty associated with $\pm$ 2 spectral subclasses.

Solar-type stars younger than 100~Myr will be contracting toward the main
sequence and will generally have lower surface gravities. Derivation of the
surface gravities and effective temperatures need to be solved jointly. First,
the effective temperature was computed assuming the star is on the main
sequence as described above. The surface gravity was then
estimated from the \citet{DM97} pre-main-sequence evolutionary tracks using the
derived temperature and assumed age from Hillenbrand \etal~(in preparation). If
the temperature was estimated from the spectral type, an iterative correction
needs to be applied since the derived temperature depends on both the spectral
type and surface gravity. For the estimated surface gravity, a new temperature
was derived using the effective temperature as a function of spectral type and
surface gravity relation in \citet{Gray92}. With the new temperature, the
surface gravity was re-derived from the \citet{DM97} evolutionary tracks.

Finally, the metallicity was fixed to $[$Fe/H$]$=0.13 for the Hyades stars 
following the measurements from \citet{Paulson03}. For all other stars, we
assumed $[$Fe/H$]$=0. The adopted metallicity, effective temperature, and
surface gravity for each star in the FEPS sample are listed in 
Table~\ref{tbl:prop}.

\section{Model Photospheres}
\label{kurucz}

In several FEPS studies, the observed \Spitzer\ flux densities were compared to
model photospheric flux densities to infer the presence of an infrared excess
diagnostic of a circumstellar disk. Model flux densities were estimated from 
synthetic photosphere spectra computed by R. 
Kurucz\footnote{http://kurucz.harvard.edu} from ATLAS 9 stellar atmospheric 
models with convective overshoot and a microturbulent velocity of 
1~km~s$^{-1}$. In this section, we describe the procedures used to normalize 
the synthetic spectra to observed photometry and to compute model flux 
densities. 

\subsection{Optical and Near-infrared Photometry}
\label{kurucz:phot}

Synthetic spectra were normalized to published optical and
near-infrared broad-band photometry. Photometric catalogs incorporated for this
study include Tycho-2 \citep{Hog00}, Hipparcos \citep{Perryman97}, 2MASS
\citep{Skrutskie06}, and the General Catalogue of Photometric Data
\citep[GCPD;][]{Mermilliod97}. The GCPD is a compilation of published ground
based observations that includes, among many others, {\it UBV} Johnson, {\it
RI} Cousins and Kron, and Stromgren {\it uvby}. The GCPD data are of
non-uniform quality compared to these other surveys.
Ground-based infrared photometry from the \ISO\ preparatory observations in
both the ESO and Tenerife photometric
systems\footnote{http://www.iso.vilspa.esa.es/users/expl\_lib/ISO/wwwcal/isoprep/gbpp/photom}
were also included. Finally, the FEPS team obtained $BRVI$ photometry for
several stars that did not have high-quality photometry available in the
literature. The observations, data reduction, and measured photometry for these
sources are presented in Appendix~\ref{newphot}.

\subsection{Synthetic Photometry}
\label{kurucz:synthetic}

For wavelengths longer than 10$\mu$m, the original Kurucz synthetic spectra are
sampled at 10.02$\mu$m, and then between 20 and 160$\mu$m in steps of 20$\mu$m.
For wavelengths longer than 10$\mu$m, we resampled the Kurucz spectra at finer
wavelengths by interpolating between model data points assuming a $S_\nu
\propto \nu^2$ spectrum.

Synthetic fluxes were computed by multiplying a Kurucz synthetic spectrum with
the spectral response of a photometric system. The spectral response,
$T(\lambda)$, includes the detector quantum efficiency, the atmospheric
transmission (if appropriate), the filter transmission, and any other optics
whose characterizations are available (see Cohen \etal~1999 for details). The
product of these three transmission functions are referred to as a FAD (i.e.
filter + atmosphere + detector).     

By definition, the bandwidth of the filter in wavelength and frequency units is
\begin{eqnarray}
     \Delta\lambda & = & \int T(\lambda) / T_{\rm max}\ d\lambda \\
     \Delta\nu     & = & \int T(\lambda) / T_{\rm max}\ d\nu,
\end{eqnarray}
where $T_{\rm max}$ is the peak transmission. Uncertainties in the 
bandwidths were computed by assuming a 5\% uncertainty in the transmission 
at any given wavelength. The spectral irradiance, $I$, can be computed by 
integrating the spectrum, $S(\lambda)$, over the FAD as
\begin{equation}
     I = \int S(\lambda)\ T(\lambda)\ d\lambda.\label{eqn:irr}
\end{equation}
The corresponding (isophotal) flux density is then defined as
\begin{eqnarray}
     S_\lambda & = & I / \Delta\lambda \\
     S_\nu     & = & I / \Delta\nu.
\end{eqnarray}

Since observed optical and near-infrared flux densities are typically quoted 
in magnitudes, the synthetic measurements were converted to magnitudes based 
on the flux for a zero-magnitude star as
\begin{equation}
\label{eq:zp}
     m = -2.5\log\Big({S_\lambda \over ZP}\Big) + zpo,
\end{equation}
where $ZP$ is the zero point of the photometric system, and $zpo$ is the offset
needed to convert the synthetic photometry to the observed photometric system.
Martin Cohen and collaborators have produced a series of papers in which they
define the zero points and zero point offsets for several photometric systems. 
We adopt the calibration by \citet{Cohen03a} for 2MASS, \citet{Cohen03b} for
Tycho-2, Hipparcos, and Landolt $BVRI$, and \citet{Cohen99} for ESO $HK$ and
Tenerife $HK$. For Stromgren photometry, we adopt the calibration of
\citet{Gray98}, but replace his flux density for Vega at 5556~\AA\ with that 
of \citet{Cohen92} for consistency.

\subsection{Fitting Procedure}
\label{kurucz:fit}

The $\chi^2$ merit equation to determine the best fit Kurucz model is
\begin{eqnarray}
    \label{eq:fit}
    \chi^2 = \sum_{i=1}^N
       \Bigl(
       {[F_{i,\rm obs} - 
         F_{i,\rm model}(T_{\rm eff},{\rm A_V,[Fe/H], log}~g, \Omega)]^2 \over 
           (\Delta F_{i,\rm obs}^2 + \Delta F_{i,{\rm model}}^2)}\Bigr)
     + \ \ \Big({T_{\rm eff} - T_{\rm eff,o} \over 
           \Delta T_{\rm eff,o}}\Big)^2,
\end{eqnarray}
where $F_{i,\rm{obs}}$ is the observed flux density typically expressed in 
magnitudes, $F_{i,{\rm model}}$ is the model flux density that depends on the 
stellar effective temperature ($T_{\rm eff}$), visual extinction (A$_{\rm V}$), 
metallicity ($\rm [Fe/H]$), surface gravity (${\rm log}~g$), and 
solid angle ($\Omega$), and $T_{\rm eff,o}$ is the nominal temperature of 
the star derived from the spectral type (if available).

Equation~\ref{eq:fit} was minimized using a modified version of the
Levenberg-Marquardt method as implemented by the LMDIF routine in the MINPACK
library\footnote{http://www.netlib.org/minpack}. The model parameters are the
solid angle of the star, the effective temperature, surface gravity,
metallicity, and visual extinction. In practice, the metallicity and surface
gravity was fixed to the values listed in Table~\ref{tbl:prop}. The constraint
in the fitting procedure is that the visual extinction is non-negative. The
initial values for ${\rm A_V}$ and $T_{\rm eff}$ were set based on the stellar
properties (see Appendix~\ref{stellar}).

Fits were constrained using photometry at wavelengths between 0.4 
and 2.5\micron\ for most sources. A few sources have excesses at $K$-band
\citep{Silverstone06} and the model was fitted to photometry between 
0.4 and 1.2\micron. Shorter wavelength photometry, in particular $U$-band 
observations, were omitted since those data are difficult to calibrate from 
the ground and are sensitive to the stellar metallicity. Longer wavelengths 
were omitted to avoid having infrared excesses bias the model fits. 

Uncertainties in the model flux densities were computed using a grid search
around the best-fit model parameters. The size of the grid was $\pm$3 times the
nominal parameter uncertainties computed from the covariance matrix computed
from the least-squares fit. At each point in the model grid, we computed
model flux densities, including the \Spitzer\ IRAC and MIPS photometric bands, 
as well as the $\chi^2$ between that model and the observed flux densities for 
photometric bands between 0.4 and 
2.5\micron. The relative probability that the model at a given grid point can 
reproduce the observations is $e^{-\chi^2/2}$. The probabilities over all
grid points then yields the probability distribution of model flux densities.

It is not feasible to present the full probability distribution for each
\Spitzer\ photometric band and each star. We instead characterized the
probability distribution for a photometric band by the nominal flux density,
$F_{\rm model}$, and the 1$\sigma$ uncertainty $\Delta F_{\rm model}$. The 
nominal flux density is given by the flux density computed from the best fit
model parameters. The 1$\sigma$ flux uncertainty is defined as the smallest
range of model flux densities about $F_{\rm model}$ that encompasses 68\% of 
the total probability. Results from the Kurucz-model fitting have been
used by \citet{Kim05}, \citet{Hines06}, and \citet{Hillenbrand08}.

\clearpage



\begin{deluxetable}{lcc@{\extracolsep{2pt}}ccccccccc@{\extracolsep{-5pt}}ccccc}
\rotate
\tabletypesize{\tiny}
\tablewidth{0pt}
\tablecaption{IRAC and MIPS Photometry\label{tbl:phot}}
\tablehead{
  \colhead{Source} &
  \multicolumn{2}{c}{IRAC 3.6\micron}  &
  \multicolumn{2}{c}{IRAC 4.5\micron}  &
  \multicolumn{2}{c}{IRAC 8\micron}  &
  \multicolumn{2}{c}{MIPS 24\micron} &
  \multicolumn{2}{c}{MIPS 70\micron} &
  \colhead{IRAC} &
  \colhead{MIPS 24\micron} &
  \multicolumn{2}{c}{AORKEY} &
  \colhead{IRAC} &
  \colhead{MIPS}\\
  \cline{2-3}
  \cline{4-5}
  \cline{6-7}
  \cline{8-9}
  \cline{10-11}
  \cline{14-15}
  \colhead{}         &
  \colhead{S$_\nu$}  &
  \colhead{$\sigma_{\rm int}$}  &
  \colhead{S$_\nu$}  &
  \colhead{$\sigma_{\rm int}$}  &
  \colhead{S$_\nu$}  &
  \colhead{$\sigma_{\rm int}$}  &
  \colhead{S$_\nu$}  &
  \colhead{$\sigma_{\rm int}$}  &
  \colhead{S$_\nu$}  &
  \colhead{$\sigma_{\rm int}$}  &
  \colhead{frame time} &
  \colhead{DCE Time} &
  \colhead{IRAC} &
  \colhead{MIPS} &
  \colhead{Flags} &
  \colhead{Flags}\\
  \colhead{} &
  \colhead{(mJy)} &
  \colhead{(mJy)} &
  \colhead{(mJy)} &
  \colhead{(mJy)} &
  \colhead{(mJy)} &
  \colhead{(mJy)} &
  \colhead{(mJy)} &
  \colhead{(mJy)} &
  \colhead{(mJy)} &
  \colhead{(mJy)} &
  \colhead{(sec)} &
  \colhead{(sec)}
}
\startdata
          1E 0307.4+1424 &     91.21 &     0.66 &    57.60 &     0.70 &    20.51 &     0.14 &    2.60 &      0.03 &   -10.8 &       5.4 &       0.40 &         9.96 &      5296896 &      5297408 &            &         \\
          1E 0324.1-2012 &     86.59 &     0.62 &    55.47 &     0.68 &    19.53 &     0.13 &    2.50 &      0.02 &    -1.7 &       3.7 &       0.40 &         9.96 &      5297664 &      5298176 &            &         \\
   1RXS J025216.9+361658 &    321.18 &     2.31 &   202.24 &     2.47 &    77.25 &     0.51 &    8.81 &      0.08 &    -0.6 &       4.3 &       0.10 &         9.96 &      5256960 &      5257472 &            &        2\\
   1RXS J025751.8+115759 &    121.01 &     0.87 &    76.26 &     0.93 &    27.93 &     0.18 &    3.27 &      0.04 &     1.3 &       5.4 &       0.10 &         9.96 &      5261568 &      5262080 &            &         \\
   1RXS J030759.1+302032 &    313.33 &     2.26 &   196.62 &     2.40 &    71.77 &     0.47 &    8.13 &      0.08 &     2.0 &       5.0 &       0.10 &         2.62 &      5313536 &      5314048 &            &         \\
   1RXS J031644.0+192259 &     55.94 &     0.40 &    35.50 &     0.43 &    12.87 &     0.09 &    1.57 &      0.03 &    -9.4 &       5.9 &       0.40 &         9.96 &      5302272 &      5302784 &            &         \\
   1RXS J031907.4+393418 &     47.01 &     0.34 &    29.19 &     0.36 &    10.88 &     0.07 &    1.31 &      0.02 &    -3.0 &       4.3 &       0.40 &         9.96 &      5257728 &      5258240 &            &         \\
   1RXS J034423.3+281224 &    389.93 &     2.81 &   246.84 &     3.01 &    88.38 &     0.58 &   10.01 &      0.09 &    -0.4 &       5.5 &       0.10 &         2.62 &      5258496 &      5259008 &            &         \\
\enddata
\tablenotetext{1}{IRAC photometry performed using aperture radius of 4 pixels}
\tablenotetext{2}{Neighboring source subtracted from MIPS 70um image before measuring photometry}
\tablenotetext{3}{IRAC photometry measured using non-FEPS, archival data}
\tablenotetext{4}{MIPS photometry measured using non-FEPS, archival data}
\end{deluxetable}




\begin{deluxetable}{cccc}
\tablewidth{0pt}
\tablecaption{IRAC Calibration Offsets Relative to 0.40 sec frame-time\label{tbl:irac_offsets}}
\tablecolumns{4}
\tablehead{
  \colhead{Band} &
  \multicolumn{3}{c}{IRAC Frame-time} \\
  \cline{2-4}
  \colhead{} &
  \colhead{0.02 sec} &
  \colhead{0.10 sec} &
  \colhead{0.60 sec}
}
\startdata
3.6\micron & $0.961 \pm 0.006$ & $1.014 \pm 0.007$ & $0.988 \pm 0.007$ \\
4.5\micron & $0.964 \pm 0.006$ & $1.021 \pm 0.007$ & $0.988 \pm 0.007$ \\
8.0\micron & $0.971 \pm 0.005$ & $1.014 \pm 0.007$ & $0.962 \pm 0.006$ \\
\enddata
\end{deluxetable}





\begin{deluxetable}{lrrrr}
\tabletypesize{\footnotesize}
\tablewidth{0pt}
\tablecaption{New $BVRI$ Photometry for FEPS Sources\label{tbl:newphot}}
\tablehead{
  \colhead{Source} &
  \colhead{$B_{\rm J}$} &
  \colhead{$V_{\rm J}$} &
  \colhead{$R_{\rm C}$} &
  \colhead{$I_{\rm C}$}
}
\startdata
1E 0307.4+1424 & 11.129 $\pm$ 0.030 & 10.463 $\pm$ 0.025 & 10.045 $\pm$ 0.023 & 9.693 $\pm$ 0.023\\
1E 0324.1-2012 & 11.049 $\pm$ 0.037 & 10.435 $\pm$ 0.031 & 10.082 $\pm$ 0.028 & 9.726 $\pm$ 0.024\\
1RXS J025216.9+361658 & 11.839 $\pm$ 0.031 & 10.715 $\pm$ 0.026 & 10.016 $\pm$ 0.019 & 9.369 $\pm$ 0.024\\
1RXS J025751.8+115759 & 11.645 $\pm$ 0.031 & 10.797 $\pm$ 0.026 & 10.266 $\pm$ 0.024 & 9.799 $\pm$ 0.024\\
1RXS J031644.0+192259 & 11.598 $\pm$ 0.030 & 10.996 $\pm$ 0.025 & 10.598 $\pm$ 0.023 & 10.248 $\pm$ 0.023\\
1RXS J031907.4+393418 & 12.444 $\pm$ 0.030 & 11.653 $\pm$ 0.025 & 11.159 $\pm$ 0.023 & 10.734 $\pm$ 0.023\\
1RXS J035028.0+163121 & 11.257 $\pm$ 0.030 & 10.567 $\pm$ 0.025 & 10.096 $\pm$ 0.023 & 9.701 $\pm$ 0.023\\
1RXS J051111.1+281353 & 11.492 $\pm$ 0.053 & 10.522 $\pm$ 0.028 & 9.933 $\pm$ 0.030 & 9.398 $\pm$ 0.028\\
\enddata
\end{deluxetable}





\begin{deluxetable}{lccr@{\extracolsep{2pt}}c@{\extracolsep{-2pt}}cr@{\extracolsep{2pt}}c@{\extracolsep{-2pt}}cr}
\tablewidth{0pt}
\tablecaption{Adopted Intrinsic Colors\label{tbl:colors}}
\tablehead{
  \colhead{SpT} &
  \multicolumn{3}{c}{$(B-V)_{\rm T}$} &
  \multicolumn{3}{c}{$J-K_s$} &
  \multicolumn{3}{c}{$V_{\rm T}-K_s$} \\
  \cline{2-4}
  \cline{5-7}
  \cline{8-10}
  \colhead{}         &
  \colhead{Average} &
  \colhead{Dispersion}  &
  \colhead{$N$}  &
  \colhead{Average} &
  \colhead{Dispersion}  &
  \colhead{$N$}  &
  \colhead{Average} &
  \colhead{Dispersion}  &
  \colhead{$N$}
}
\startdata
F2 & 0.383 & 0.028 &  74 & 0.224 & 0.029 & 153 & 0.980 & 0.093 &  58\\
F3 & 0.433 & 0.025 & 147 & 0.237 & 0.019 & 257 & 1.090 & 0.057 & 102\\
F5 & 0.483 & 0.022 & 266 & 0.267 & 0.026 & 491 & 1.216 & 0.069 & 230\\
F6 & 0.527 & 0.022 & 214 & 0.288 & 0.020 & 337 & 1.305 & 0.058 & 173\\
F7 & 0.558 & 0.026 & 250 & 0.303 & 0.025 & 371 & 1.354 & 0.050 & 177\\
F8 & 0.592 & 0.023 & 130 & 0.320 & 0.025 & 211 & 1.393 & 0.043 &  90\\
G0 & 0.632 & 0.025 & 206 & 0.329 & 0.021 & 313 & 1.460 & 0.041 & 145\\
G1 & 0.648 & 0.023 & 114 & 0.342 & 0.025 & 197 & 1.498 & 0.067 & 117\\
\enddata
\end{deluxetable}





\begin{deluxetable}{lcc@{\extracolsep{2pt}}cc@{\extracolsep{2pt}}cc@{\extracolsep{2pt}}cc@{\extracolsep{2pt}}cc@{\extracolsep{2pt}}c@{\extracolsep{-5pt}}ccccc}
\tabletypesize{\small}
\tablewidth{0pt}
\tablecaption{Adopted Stelllar Properties\label{tbl:prop}}
\tablehead{
  \colhead{Source} &
  \colhead{SpT} &
  \colhead{${\rm A_V}$} &
  \colhead{${\rm[Fe/H]}$} &
  \colhead{${\rm log}\ g$} &
  \colhead{$T_{\rm eff}$} \\
  \colhead{} &
  \colhead{} &
  \colhead{(mag)} &
  \colhead{(dex)} &
  \colhead{(log g cm$^{-2}$)} &
  \colhead{(K)}
}
\startdata
1E 0307.4+1424          & G6V        & 0.03       & 0.00       & 4.54       & 5591      \\
1E 0324.1-2012          & G4V        & 0.04       & 0.00       & 4.54       & 5745      \\
1RXS J025216.9+361658   & K2IV       & 0.95       & 0.00       & 4.63       & 4999      \\
1RXS J025751.8+115759   & G7V        & 0.62       & 0.00       & 4.54       & 5552      \\
1RXS J030759.1+302032   & G5IV       & 0.16       & 0.00       & 4.50       & 5676      \\
1RXS J031644.0+192259   & G2V        & 0.10       & 0.00       & 4.47       & 5886      \\
1RXS J031907.4+393418   & K0V        & 0.19       & 0.00       & 4.61       & 5287      \\
1RXS J034423.3+281224   & G7V        & 0.05       & 0.00       & 4.54       & 5725      \\
\enddata
\end{deluxetable}


\clearpage

\begin{figure}
\includegraphics[angle=-90,scale=0.9]{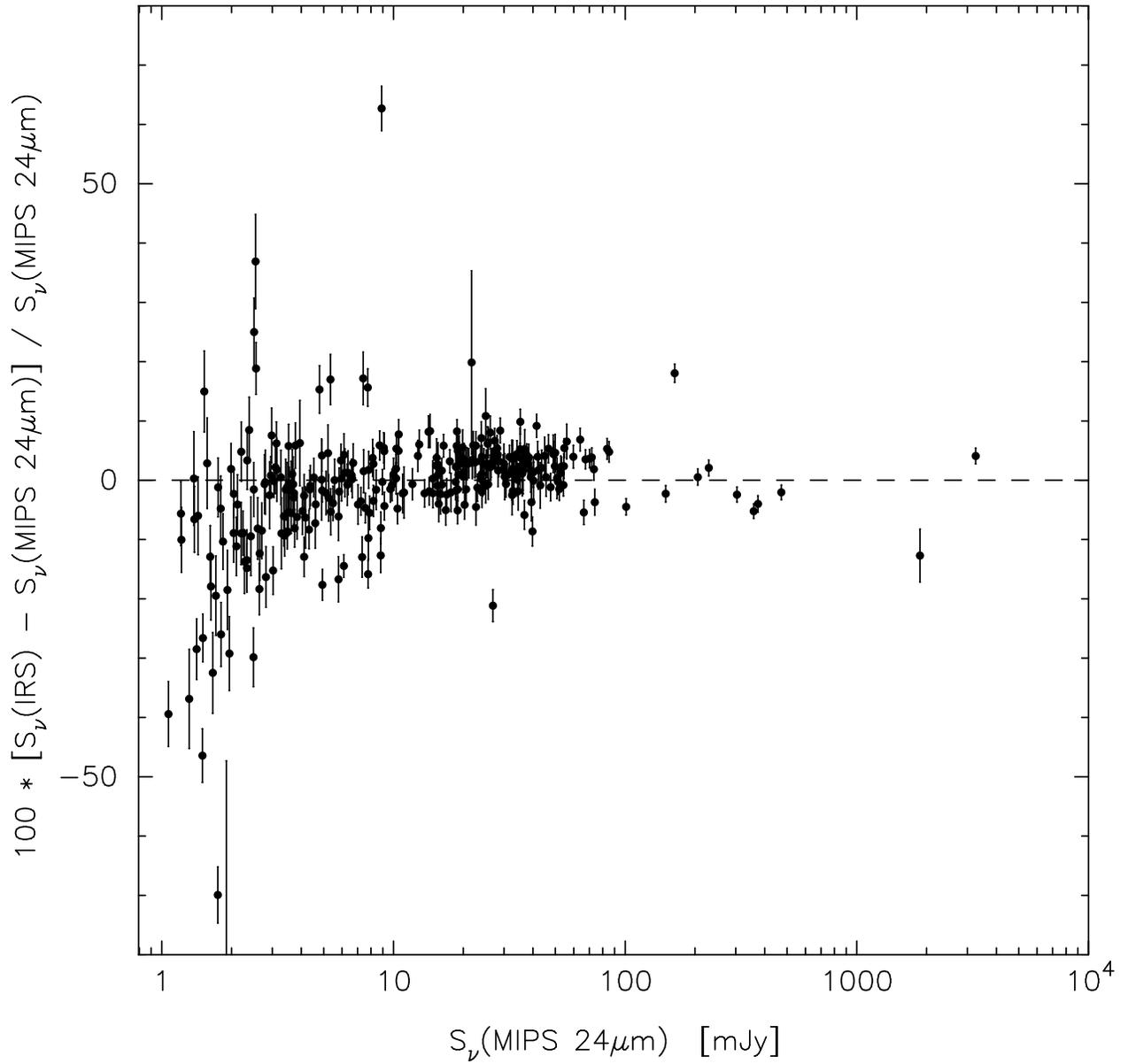}
\caption{
  \label{fig:irs_mips24}
  Percent difference between the synthetic IRS~24\micron\ photometry and
  MIPS~24\micron\ photometry as a function of the MIPS~24\micron\ flux 
  density. The IRS~24\micron\ photometry was computed by integrating the 
  observed IRS spectrum over the MIPS~24\micron\ bandpass. The horizontal
  dashed line is shown for reference.
}
\end{figure}


\clearpage

\begin{thebibliography}{}

\bibitem[Ardila \etal(2005)]{Ardila05} Ardila, D. R. \etal~2005, 
   \apjl, 617, L147
\bibitem[Bouwman \etal(2008)]{Bouwman08} Bouwman, J., \etal~2008, \apj, in press
\bibitem[Breger(1986)]{Breger86} Breger, M. 1986, \apj, 309, 311
\bibitem[Carpenter \etal(2005)]{Carpenter05} Carpenter, J. M., Wolf, S.,
   Schreyer, K., Launhardt, R., \& Henning, T. 2005, \aj, 129, 1049
\bibitem[Carpenter \etal(2008)]{Carpenter08} Carpenter, J. M., \etal~2008,
   \apj, submitted
\bibitem[Chen \etal(1998)]{Chen98} Chen, B., Vergely, J. L., Valette, B.,
   \& Carraro, G. 1998, \aap, 336, 137
\bibitem[Cohen \etal(1992)]{Cohen92} Cohen, M. Walker, R. G., Barlow, M. J., 
   \&  Deacon, J. R. 1992, \aj, 104, 1650
\bibitem[Cohen \etal(2003a)]{Cohen03a} Cohen, M., Megeath, S. T., 
   Hammersley, P. L., Mart\'in-Luis, F., \& Stauffer, J. 2003a, \aj, 125, 2645
\bibitem[Cohen \etal(1999)]{Cohen99} Cohen, M., Walker, R. G., Carter, B., 
   Hammersley, P. L., Kidger, M., \& Noguchi, K. 1999, \aj, 117, 1864
\bibitem[Cohen, Wheaton, \& Megeath(2003b)]{Cohen03b} Cohen, M., 
   Wheaton, W. A., \& Megeath, S. T. 2003b, \aj, 126, 1090
\bibitem[Crawford \& Barnes(1974)]{Crawford74} Crawford, D., \& Barnes, J. 
   1974, \aj, 79, 687
\bibitem[D'Antona \& Mazzitelli(1997)]{DM97} D'Antona, F., \&
   Mazzitelli, I. 1997, MmSAI, 68, 807
\bibitem[Engelbracht \etal(2007)]{Engelbracht07} Engelbracht \etal~2007,
   \pasp, 119, 994
\bibitem[Evans \etal(2003)]{Evans03} Evans, N. J., II, \etal~2003, 
   \pasp, 115, 965
\bibitem[Fazio \etal(2004)]{Fazio04} Fazio, G., Hora, J. L, Allen, L. E.,
    \etal~2004, \apjs, 154, 10
\bibitem[Gordon \etal(2007)]{Gordon07} Gordon, K. D., \etal~2007, 
   \pasp, 119, 1019
\bibitem[Gray(1992)]{Gray92} Gray, D. F. 1992, The Observation and Analysis 
    of Stellar Photospheres (Cambridge: Cambridge Univ. Press)
\bibitem[Gray(1998)]{Gray98} Gray, R.O. 1998, \aj, 116, 482
\bibitem[Hampel(1974)]{Hampel74} Hampel, F. 1974, J. AM. Statist. Assoc.,
   69, 383
\bibitem[Hillenbrand \etal(2008)]{Hillenbrand08} Hillenbrand, L. A., 
   \etal~2008, \apj, 677, 630
\bibitem[Hines \etal(2006)]{Hines06} Hines, D. C., \etal~2006,
   \apj, 638, 1070
\bibitem[Hollenbach \etal(2005)]{Hollenbach05} Hollenbach, D., \etal~2005,
     \apj, 631, 1180
\bibitem[H\o g \etal(2000)]{Hog00} H\o g, E., Fabricius, C., Makarov, V.~V.,
     Urban, S., Corbin, T., Wycoff, G., Bastian, U.,  Schwekendiek, P. \&
	 Wicenec, A. 2000, \aap, 355, L27
\bibitem[Houck \etal(2004)]{Houck04} Houck, J., \etal~2004, \apjs, 154, 18
\bibitem[Houdashelt, Bell, \& Sweigart(2000)]{Houdashelt00} Houdashelt, 
     M. L., Bell, R. A., \& Sweigart, A. V. 2000, \aj, 119, 1448
\bibitem[Kim \etal(2005)]{Kim05} Kim, J. S., \etal~2005, \apj, 632, 659
\bibitem[Landolt(1992)]{Landolt92} Landolt, A. U. 1992, \aj, 104, 340
\bibitem[Makovoz \& Marleau(2005)]{Makovoz05} Makovoz, D., \& Marleau, F. R.
     2005, \pasp, 117, 1113
\bibitem[Mamajek, Meyer, \& Liebert(2002)]{Mamajek02} Mamajek, E. E.,
     Meyer, M. R., \& Liebert, J. 2002, \aj, 124, 1670
\bibitem[Mamajek, Meyer, \& Liebert(2006)]{Mamajek06} Mamajek, E. E.,
     Meyer, M. R., \& Liebert, J. 2006, \aj, 131, 2360
\bibitem[Mathis(1990)]{Mathis90} Mathis, J. S. 1990, \araa, 27, 37
\bibitem[Mermilliod, Mermilliod, \& Hauck(1997)]{Mermilliod97}
     Mermilliod, J.-C., Mermilliod, M., \& Hauck, B. 1997, \aaps, 124, 349
\bibitem[Meyer \etal(2004)]{Meyer04} Meyer, M. R., \etal~2004, \apjs, 154, 422
\bibitem[Meyer \etal(2006)]{Meyer06} Meyer, M. R., \etal~2006, \pasp, 118, 1690
\bibitem[Meyer \etal(2008)]{Meyer08} Meyer, M. R., \etal~2008, \apjl, 673, L181
\bibitem[Moro-Mart\'in \etal(2007)]{Moro07} Moro-Mart\'in, A., \etal~2007,
   \apj, 658, 1312
\bibitem[Nagao \& Matsuyama(1979)]{Nagao79} Nagao, M., \& Matsuyama, T.
   1979, Edge Preserving Smoothing, Computer Graphics and Image Processing,
   vol. 9, No. 4, 394
\bibitem[Naylor(1998)]{Naylor98} Naylor, T. 1998, \mnras, 296, 339
\bibitem[Pascucci \etal(2006)]{Pascucci06} Pascucci, I. \etal~2006, \apj, 
     651, 1177
\bibitem[Pascucci \etal(2007)]{Pascucci07} Pascucci, I. \etal~2007, 
   \apj, 663, 383
\bibitem[Paulson, Sneden, \& Cochran(2003)]{Paulson03} Paulson, D. B.,
   Sneden, C., Cochran, W. D. 2003, \aj, 125, 3185
\bibitem[Perryman \etal(1997)]{Perryman97} Perryman, M. A. C., \etal~1997,
   \aap, 323, L49
\bibitem[Pinsonneault \etal(1998)]{Pinsonneault88} Pinsonneault, M. H.,
   Stauffer, J., Soderblom, D. R., King, J. R., \& Hanson, R. B. 1998,
   \apj, 504, 170
\bibitem[Prosser(1994)]{Prosser94} Prosser, C. F. 1994, \aj, 103, 488
\bibitem[Reach \etal(2005)]{Reach05} Reach, W. T., \etal~2005, \pasp, 117, 978
\bibitem[Rieke \etal(2004)]{Rieke04} Rieke, G. \etal~2004, \apjs, 154, 25
\bibitem[Rieke \etal(2008)]{Rieke08} Rieke, G. \etal~2008, \aj, 135, 2245
\bibitem[Sandage(1972)]{Sandage72} Sandage, A. 1972, \apj, 178, 1
\bibitem[Silverstone \etal(2006)]{Silverstone06} Silverstone, M. D., 
   \etal~2006, \apj, 639, 1138
\bibitem[Skrutskie \etal(2006)]{Skrutskie06} Skrutskie, M. F., \etal~2006,
   \aj, 131, 1163
\bibitem[Stauffer \etal(2005)]{Stauffer05} Stauffer, J. R., \etal~2005, 
   \aj, 130, 1834
\bibitem[Taylor(2006)]{Taylor06} Taylor, B. J. 2006, \aj, 132, 2453
\bibitem[Welsh, Crifo, \& Lallement(1998)]{Welsh98} Welsh, B. Y., Crifo, F., 
   \& Lallement, R. 1998, \aap, 333, 101
\bibitem[Werner \etal(2004)]{Werner04} Werner, M., \etal~2004, \apjs, 154, 1
\bibitem[Whiteoak(1961)]{Whiteoak61} Whiteoak, J. B. 1961, \mnras, 123, 245
\bibitem[Williams \etal(2004)]{Williams04} Williams, J. P., Najita, J., 
   Liu, M. C., Bottinelli, S., Carpenter, J. M., Hillenbrand, L. A., 
   Meyer, M. R., \& Soderblom, D. R. 2004, \apj, 604, 414
\bibitem[Wright \etal(2003)]{Wright03} Wright, C. O., Egan, M. P.,
    Kraemer, K. E., \& Price, S. D. 2003, \aj, 125, 349
\bibitem[Zacharias \etal(2004)]{Zacharias04} Zacharias, N., Urban, S. E., 
   Zacharias, M. I., Wycoff, G. L., Hall, D. M., Monet, D. G., \& 
   Rafferty, T. J. 2004, \aj, 127, 3043
\end{thebibliography}
\end{document}